\documentclass[aps,reprint,superscriptaddress]{revtex4-1}

\usepackage{graphicx}
\usepackage{physics}
\usepackage{braket}
\usepackage{hyperref}
\usepackage{amsfonts}
\usepackage{amssymb}
\usepackage{dsfont}
\usepackage{xcolor}
\usepackage{enumitem}
\usepackage{soul}
\graphicspath{{"../figures/"}}

\usepackage[normalem]{ulem}
\definecolor{DarkRed}{rgb}{0.80,0,0}
\definecolor{Orange}{rgb}{0.91,0.63,0}
\definecolor{SkyBlue}{rgb}{0.34,0.71,0.92}

\begin{document}

\title{A many-body singlet prepared by a central spin qubit
}
\author{Leon Zaporski\textsuperscript{1,*}}
\author{Stijn R. de Wit\textsuperscript{1,2,*}}
\author{Takuya Isogawa\textsuperscript{1}}
\author{Martin Hayhurst Appel\textsuperscript{1}}
\author{Claire Le Gall\textsuperscript{1}}
\author{Mete Atat\"ure\textsuperscript{1,$\dagger$}}
\author{Dorian A. Gangloff\textsuperscript{3,$\dagger$}}

\noaffiliation

\affiliation{Cavendish Laboratory, University of Cambridge, JJ Thomson Avenue, Cambridge, CB3 0HE, United Kingdom}
\affiliation{MESA+ Institute for Nanotechnology, University of Twente, The Netherlands}
\affiliation{Department of Engineering Science, University of Oxford, Parks Road, Oxford, OX1 3PJ, United Kingdom
\\ \ \\
\textsuperscript{*}\,These authors contributed equally\\
\textsuperscript{$\dagger$}\,Correspondence to: ma424@cam.ac.uk, dorian.gangloff@eng.ox.ac.uk.
\\ \ \\
}

\begin{abstract}
Controllable quantum many-body systems are platforms for fundamental investigations into the nature of entanglement and promise to deliver computational speed-up for a broad class of algorithms and simulations. In particular, engineering entanglement within a dense spin ensemble can turn it into a robust quantum memory or a computational platform. Recent experimental progress in dense central spin systems motivates the design of algorithms that 
use a central-spin qubit as a convenient proxy for the ensemble. Here we propose a protocol that uses a central spin to initialize two dense spin ensembles into a pure anti-polarized state and from there
creates a many-body entangled state -- a singlet -- from the combined ensemble. We quantify the protocol performance for multiple material platforms and show that it can be implemented even in the presence of realistic levels of decoherence. Our protocol introduces an algorithmic approach to preparation of a known many-body state and to entanglement engineering in a dense spin ensemble, which can be extended towards a broad class of collective quantum states. 
\end{abstract}
\maketitle

\section{Introduction}

Controlling quantum properties in many-particle systems, whether for technological advantage or foundational studies, can be reduced to controlling the relative participation and phase of the system's eigenstates. In most cases, this leads to entanglement amongst the system's particles \cite{Takou2023}. The initialization of a quantum system to a pure state is a necessary starting point from which to engineer entanglement \cite{DiVincenzo2000a}. Initialization through traditional cooling techniques brings a quantum system in contact with a bath whose temperature is below that of the system's characteristic energy scale, bringing the target system to its ground state \cite{Wineland2013}. An equivalent picture exists for driven systems, where directionality within an energy hierarchy of dressed states can bring the system to an effective ground state of the dressed system \cite{Kessler2012}. The latter approach is more versatile as it allows the design of tailored ground states. Crucially, it also requires access to the degree of freedom one intends to cool. Versions of this approach appear in driven-dissipative state preparations of photonic systems \cite{Bardyn2012,Marino2019}, diamond color centers \cite{Chen2017,Greiner2017,Ikeda2020}, epitaxial quantum dots \cite{Issler2010,Ethier-Majcher2017}, and trapped atoms in an optical resonator \cite{Zhu2019}.

Central-spin systems, typically consisting of a single electronic spin coupled to an ensemble of nuclear spins \cite{Urbaszek2013}, have been a particularly potent testing ground for the active approaches to state preparation \cite{Bluhm2011,Jackson2022}. Firstly, this is by necessity: nuclear-spin energy scales make ground state preparation well beyond the reach of modern refrigeration techniques. Secondly, the one-to-all coupling of the central spin qubit to an ensemble of spins yields a convenient proxy for the ensemble spins \cite{Jackson2021}. In the few-spin regime of diamond color centers or rare earth ions, dynamic nuclear polarization can initialize a set of proximal spins to a high-purity polarized state \cite{Schwartz2018}. This is then the starting point for qubit storage in an ensemble excitation \cite{Ruskuc2022}, two-body singlet engineering \cite{Bartling2022}, or spin-by-spin quantum computation \cite{Abobeih2022}.

In the limit of dense ensembles, the constituent spins are indistinguishable when interacting with the central spin. Entanglement is then readily generated and measured by controlling the central spin's dynamics\cite{Gangloff2019,Jackson2021,Gangloff2021}. Collective states of the ensemble become natural targets of a driven purification technique\cite{Jackson2022}. In addition, an effective all-to-all coupling mediated by the central spin \cite{Wust2016} leads to a highly-correlated behavior of the ensemble which, in principle, can be harnessed for state engineering. Despite the absence of individual spin control, the dense systems of interest\cite{Zaporski2022,Stockill2016,Bluhm2011} operate in a truly many-body regime of $10^{4}$ to $10^{6}$ interacting ensemble spins - far exceeding the number of physical qubits present in near-term quantum simulators \cite{Taminiau2014,Browaeys2020}.

A many-body singlet is a superposition of ensemble spins with zero angular momentum. The singlet state is a hyperfine vacuum and renders the bath invisible to the central spin, protecting the latter from interacting with its environment\cite{Chen2017,PhysRevX.6.021040,PhysRevLett.111.173002} -- a decoherence-free subspace for the central spin qubit. Owing to destructive interference between pairs of ensemble spins, the ensemble is also protected from noise whose length scale is larger than the system size. This makes it a particularly useful and readily detectable first target state to prepare in a central-spin system. Further, many-body singlet preparation offers insights into the evolution of entanglement under slower intra-bath interactions, as well as anomalous spin diffusion at longer time scales \cite{Zu2021}.

To this date, state engineering efforts in dense central spin systems remain confined to tuning the mean field degrees of freedom, such as ensemble polarization and its fluctuations\cite{Hogele2012,Gangloff2019,YangSham2013,Jackson2022}, and classical correlations amongst ensemble spins\cite{Gangloff2021}. A degree of purification has been achieved via polarization of 80$\%$  through optical techniques\cite{Chekhovich2017} and nearly 50$\%$ via central-spin control\cite{Gangloff2021} -- such approaches proving to be generally challenging due to a dependence of the central spin transition frequency on the polarization of the bath. Meanwhile stabilizing the ensemble polarization via the central spin \cite{Jackson2022} can approach the quantum limit of polarization stability. Despite these efforts, the resulting nuclear states remain highly mixed, featuring little inter-particle phase coherence. A degree of quantum correlation among spins was observed when probing a partially polarized ensemble via the central spin \cite{Gangloff2021}, which suggested the possibility of purification via reduced total angular momentum states, so-called dark states \cite{Taylor2003}. Theoretical proposals to engineer the state of dense central spin systems have focused on dissipative phase transitions\cite{Kessler2012}, or quantum memory\cite{Taylor2003, Issler2010} and spin squeezing schemes\cite{Rudner2011} relying on fully polarized initial states. A protocol for direct control over the ensemble's inter-particle phase, allowing for purification and entanglement engineering at low total ensemble polarization, is still missing. 

In this work, we use a simple and realistic form of symmetry breaking to gain control over an ensemble's inter-particle phase. Leveraging this control, we propose a three-stage state engineering protocol that utilizes the effective interaction between two distinct spin species, naturally present in real physical systems\cite{Zaporski2022,Bluhm2011,Stockill2016}, and proceeds via control of the central spin exclusively. The first stage of the protocol locks the total polarization of the system to zero\cite{Jackson2022}. The second stage initializes the two spin species to an anti-polarized state with near-unit purity. The third stage involves a sequence of unitary gates that drives the system into a many-body singlet via phase engineering. Having in mind a near-term experimental realization, we quantify the protocol robustness as a function of model parameters and identify candidate physical platforms in which it could be successfully implemented.

\section{Main}

\begin{figure*}
    \centering
    \includegraphics[width=\textwidth]{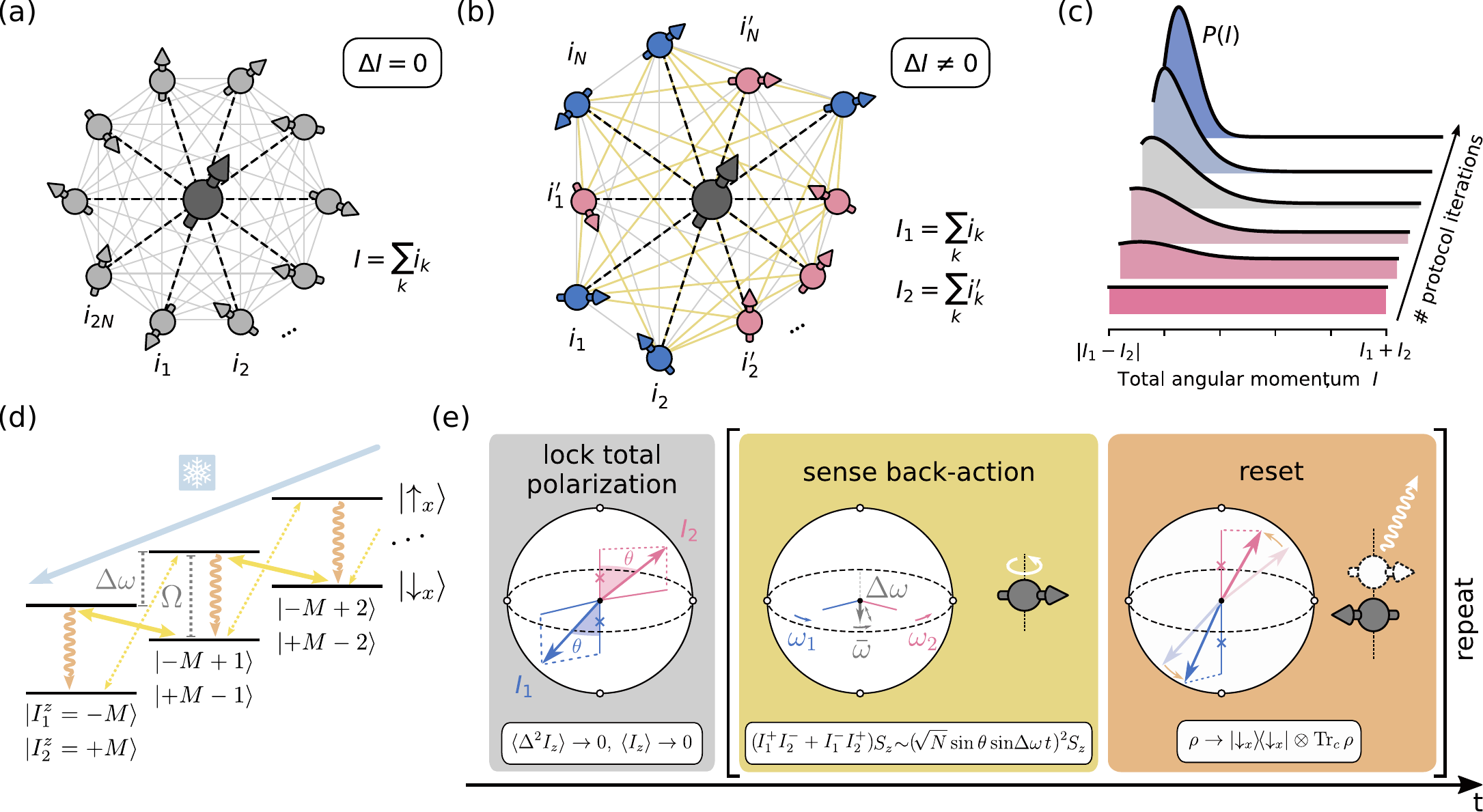}
    \caption{\textbf{State engineering in the central spin system} \textbf{a,} Perfectly homogeneous central spin system, symmetry-protected from changing its total angular momentum.
    \textbf{b,} Symmetry-broken central spin system,
    consisting of spin species $I_1$ and $I_2$,
    no longer protected from changing its total angular momentum.
    \textbf{c,} Reduction of the total angular momentum magnitude via the state purification protocol, resulting in a narrowed steady-state distribution (blue-shaded curve), peaked around $I=N^{-1/4}$. 
    \textbf{d,} Effective Jaynes-Cummings ladder of states $\ket{I_1^z=-M+n,I_2^z=M-n}$, with anti-correlated $I_1^z$ and $I_2^z$, where $M=\min(I_1,I_2)$. The thick and faint yellow arrows illustrate the dominant, and $2\Delta \omega$-detuned three-body interactions, respectively. The curvy orange arrows represent the central spin resets. The pale blue arrow displays a net direction of the phase space flow within this effective sideband-cooling process.
    \textbf{e,}
    Stages of anti-polarized state preparation.
    The red and blue arrows within the generalized Bloch spheres denote the total angular momenta of the two species, $\mathbf{I}_1$ and $\mathbf{I}_2$. Stage one (left panel): Locking of the total polarization, $I^z=I_1^z+I_2^z$, to zero. Stage two: activation of the three-body interaction (middle panel), equivalent to the back-action sensing, followed by central spin reset (right panel). 
   } 
    \label{fig:my_label}
\end{figure*}

\subsection{Symmetry and the total angular momentum representation}

Central spin systems feature high-dimensional Hilbert spaces. In the dense limit, the central spin's interaction with the ensemble does not distinguish individual spins. This leads to collective symmetries and corresponding constants of motion, which we focus on here.

In this simplest scenario of a perfectly homogeneous spin bath (see Fig.~\ref{fig:my_label}a), a general system Hamiltonian is unchanged under the re-ordering of ensemble-spin indices. Such invariance is the highest symmetry that a spin bath can exhibit, and it results in the emergence of a constant of motion that dictates the rate of collective dynamics, as in Dicke superradiance\cite{Dicke93}. Within the bath of spin-$1/2$ particles, this constant is identical to the magnitude of the total angular momentum, $I$, directly related to the eigenvalue of:
    \begin{equation}
        \Big(\sum_{k=1}^{N} \mathbf{i}_k\Big)^2 \equiv \mathbf{I}^2,
    \end{equation}
where $\mathbf{i}_k$ is the single spin operator of the $k$-th spin in the ensemble, and $N$ is the total number of ensemble spins. Ensembles of higher-spin particles, such as spin-$3/2$, can be described by equivalent, albeit more numerous, constants of motion.

Under this symmetry, coupling to the central spin cannot change the magnitude of the total angular momentum, only the polarization, and coherent control over the ensemble is thereby limited. In particular, efforts to prepare a many-body singlet ($I=0$) are futile and the ensemble dynamics are governed by thermally (with $\beta \sim 0$) occupied states of $I$, dominated by the highest degeneracy states near $I \sim \sqrt{N}$. 

We propose making use of the simplest form of reduced symmetry to gain control: breaking the system into two distinguishable but equally abundant ensembles (see Fig.~\ref{fig:my_label}b), which we call species. The groups of spins of the same species are characterized by their individual total angular momenta:  $\mathbf{I}_1$ and $\mathbf{I}_2$. Their magnitudes, $I_1$ and $I_2$, become the new constants of motion. In the discussions to follow, we assign them the value $\sqrt{N/2}$, which is the most likely value for a fully mixed initial state. Our results hold for all $(I_1,I_2)$ values in a $\sim\sqrt{N/2}$ vicinity \cite{supplementary}, thus capturing all experimentally relevant dynamics. We also note that this easily extends to more than two species.

This situation of two distinguishable spin species makes it possible to alter the magnitude of the total angular momentum, $I=|\mathbf{I}_1+\mathbf{I}_2|$, and therefore to significantly reduce it (see Fig.~\ref{fig:my_label}c). When controlled via a directional pumping process (i.e. cooling), the spin ensembles can be initialized into a pure collective anti-polarized state \cite{Sun2017}:
\begin{equation}\label{eq:classical_anti-polarized_state}
\ket{I_1^z=-M,I_2^z=M},  
\end{equation}
with $M=\min(I_1,I_2)$. Such a state contains no coherence between the species, and represents a classical limit to the total angular momentum reduction, featuring a noise of $\langle \mathbf{I}^2 \rangle_{\text{cl}}=(I_1+I_2)(|I_1-I_2|+1) \sim \sqrt{N}$; a factor $\sqrt{N}$ lower than that of a thermal state.

Creating entanglement via a controlled phase between the two species can lower the magnitude of the total angular momentum further down to $I=|I_1-I_2|$. In particular, for $I_1=I_2$,
the quantum limit is reached after the preparation of a many-body singlet:
\begin{equation}\label{eq:singlet-super position}
        \ket{I=0}=\sum_{n=0}^{2M} \frac{(-1)^n}{\sqrt{2M+1}} \ket{I_1^z=-M+n,I_2^z=M-n},
\end{equation}
expressed uniquely in the $\ket{I_1^z=-M+n,I_2^z=M-n}$-basis using Clebsh-Gordan coefficients. The many-body singlet is characterized also by a full noise suppression: $\langle \mathbf{I}^2 \rangle_{\text{qu}}=0$.

\subsection{System Hamiltonian: control via the central spin}

We take the general Hamiltonian for a dense central spin system in an external magnetic field \cite{Urbaszek2013}, and split the spin ensemble into two spin species $i=1,2$:
\begin{equation}\label{eq:Hamiltonian_CSS}
H= \omega_{\text{c}} S_z +\sum_{i=1,2} \omega_i I_i^z + \sum_{i=1,2} a_i \mathbf{S}\cdot \mathbf{I}_i.
\end{equation}
We consider the regime of high magnetic fields, as defined by a dominant Zeeman interaction $\propto \omega_{\text{c}}$ of the central spin, $\mathbf{S}$. The second term captures the internal energy structure $\omega_i$ of the spin ensembles. The last term represents a Heisenberg-type hyperfine coupling ($\propto a_i < \omega_i$) between the central spin and the bath spins. We consider a symmetry breaking between the two spin ensembles $\omega_1 \neq \omega_2$, which in real systems can take myriad forms. Without loss of generality\cite{PhysRevLett.91.246802}, we assume $a_1=a_2\equiv a$ and $\omega_1 > \omega_2$. In this high field regime, the central spin quantization axis is pinned to the $z$-direction, and the hyperfine interaction reduces to\cite{PhysRevB.77.125329,Cywinski2009}:
\begin{equation}\label{eqn:hyperfine_inter}
\begin{split}
a\sum_{i=1,2}\mathbf{S}\cdot\mathbf{I}_i &= aS_z(I_1^z+I_2^z)\\&-\frac{a^2}{4\omega_{\text{c}}}(I_1^z+I_2^z) \\&+\sum_{i=1,2} \frac{a^2}{4\omega_{\text{c}}} S_z(I_i^+ I_i^-+I_i^- I_i^+)\\
&+\frac{a^2}{2\omega_{\text{c}}} S_z(I_1^+ I_2^-+I_1^- I_2^+)\\&+ \mathcal{O}[(a/\omega_{\text{c}})^2].
\end{split}
\end{equation}
The leading and only first-order hyperfine term is the collinear interaction (used in total polarization locking). The second term renormalizes the nuclear Zeeman interaction of the ensemble by a negligible amount of $-a^2/4\omega_{\text{c}}\ll \omega_i$. 
The third and fourth terms contain the effective two-body and the three-body interactions, respectively. The former will remain off-resonance during our protocol, while the latter will be critical in cooling and correlating the two spin ensembles.

Exclusively to stabilize the ensemble polarization \cite{Gangloff2019}, we also consider that, apart from the Zeeman interaction ($\propto\omega_i$), the ensemble spins are subject to a small eigenstate-mixing interaction ($\propto \nu_i\ll \omega_i$) \cite{Hogele2012}, which results in an effective non-collinear term:
\begin{equation}\label{eqn:non_collinear}
H_\text{nc} = \sum_{i=1,2}\frac{a \nu_i}{2\omega_i}S_zI_i^x +\mathcal{O}[(\nu_i/\omega_i)^2].
\end{equation}

Finally, we consider a control of the entire system exclusively via the central spin, represented by a resonant drive of the central spin with strength $\Omega \ll \omega_{\text{c}}$, and work in a rotating frame of reference, under Rotating Wave Approximation \cite{supplementary}. 

\subsection{Preparation of a pure anti-polarized state}

Prior to the protocol execution, the spin ensemble is found in a fully-mixed state characterized by a density matrix $\rho \propto \mathds{1}$. The first two stages of our protocol initialize the ensemble to a pure collective state -- an anti-polarized state of the two spin species (Eq.~\ref{eq:classical_anti-polarized_state}) -- which enables the generation of a many-body singlet of the whole ensemble in the third stage (last section of this article).

\emph{Stage 1: polarization locking -- } The initial state exhibits maximal uncertainty of the total polarization of the bath $\langle\Delta^2 I^z\rangle\sim \sqrt{N}$, where $I^z=I_1^z+I_2^z$. The first stage of the protocol reduces this uncertainty to zero using a previously developed technique \cite{Jackson2022}. In brief terms, the collinear hyperfine interaction (the first term in Eq.~\ref{eqn:hyperfine_inter}) allows the central spin to sense the polarization-deviation from the $I^z=0$ lock-point which induces a Larmor precession around the $z$-axis. Subsequently, this deviation is corrected by a resonantly activated ($\Omega=(\omega_1+\omega_2)/2$ - c.f. Ref. \cite{HENSTRA1988389}) non-collinear interaction (using Eq.~\ref{eqn:non_collinear}) which translates the acquired phase into a change in the total ensemble polarization. Finally, the state of the central spin is reset and this stage is repeated until the ensemble reaches the limit of $\langle I^z \rangle =0$ and $\langle\Delta^2 I^z\rangle=0$. From this point on, the non-collinear interaction remains off-resonant.

\emph{Stage 2: full purification -- } At the end of the first stage, the ensemble's locked zero-polarization state is a mixture of states $\ket{I_1^z=-M+n, I_2^z=M-n}$ where $n=0,1,..,2M$, for which the remaining uncertainty lies in the polarization of individual species, $I_1^z$ and $I_2^z$. The second stage of the protocol removes this uncertainty to produce a pure collective state. Doing so relies on the driven resonant activation $\Omega= \Delta \omega \equiv \omega_1-\omega_2$ of the three-body interaction (fourth term of Eq.~\ref{eqn:hyperfine_inter}). For the central spin initialized in state $\ket{\downarrow_x}$, this interaction activates the transition:

\begin{equation}\label{eqn:three-body-exchange}
\begin{split}
&\ket{I_1^z=-M+n, I_2^z=M-n}\ket{\downarrow_x}\\
&\longleftrightarrow\ket{I_1^z=-M+n-1, I_2^z=M-n+1}\ket{\uparrow_x}.
\end{split}
\end{equation}

This can be visualized in an effective Jaynes-Cummings ladder of states (Fig.~\ref{fig:my_label}d) parameterized by the principal quantum number $n$ corresponding to the $\ket{I_1^z=-M+n, I_2^z=M-n}$ state. Equation~\ref{eqn:three-body-exchange} then represents a single quantum $n \leftrightarrow n-1$ transition down the ladder. Combined with a central spin reset $\ket{\uparrow_x} \rightarrow \ket{\downarrow_x}$ applied every half-period of the three-body interaction, repeating the transition and reset forms a directional pumping process -- equivalent to sideband cooling in harmonic systems \cite{doi:10.1126/science.aan5614} -- towards the ladder's ground state $n=0$. This ground state consists of the fully anti-polarized ensemble state $\ket{I_1^z=-M,I_2^z=M}$, which is a pure collective state of the ensemble.

Maintaining the directionality of the pumping process relies on a detuning of $2 \Delta \omega$ between the selected transition $\ket{I_1^z, I_2^z}\ket{\downarrow_x} \leftrightarrow \ket{I_1^z - 1, I_2^z + 1}\ket{\uparrow_x}$ and the unwanted transition $\ket{I_1^z, I_2^z}\ket{\downarrow_x}\leftrightarrow\ket{I_1^z + 1, I_2^z - 1}\ket{\uparrow_x}$ (thick and faint yellow arrows in Fig.~\ref{fig:my_label}d, respectively), which drive the system towards the opposite ends of the ladder. This is sustained as long as the three-body interaction strength, $Na^2/2\omega_{\text{c}}$, remains smaller than $\Delta \omega$. We characterize the directionality with their ratio, which ultimately sets the purification limit:

\begin{equation}\label{eqn:kappa}
\kappa =\omega_{\text{c}} \Delta \omega/(Na^2).
\end{equation}

We summarize this two-stage cooling process in Fig.~\ref{fig:my_label}e, where the initial locked state has two spin ensembles with arbitrary but opposite orientations. Semi-classically, the second stage (three-body interaction) is equivalent to the central spin sensing the Larmor precession beatnote between the two spin ensembles (recall $\omega_1 \neq \omega_2$) in the hyperfine back-action, which acts on the central spin akin to a classical driving field. The timing of the central spin reset favors a particular phase of this beatnote, and thus prepares the two ensembles in a specific orientation.

\begin{figure}
    \centering
    \includegraphics[width=\columnwidth]{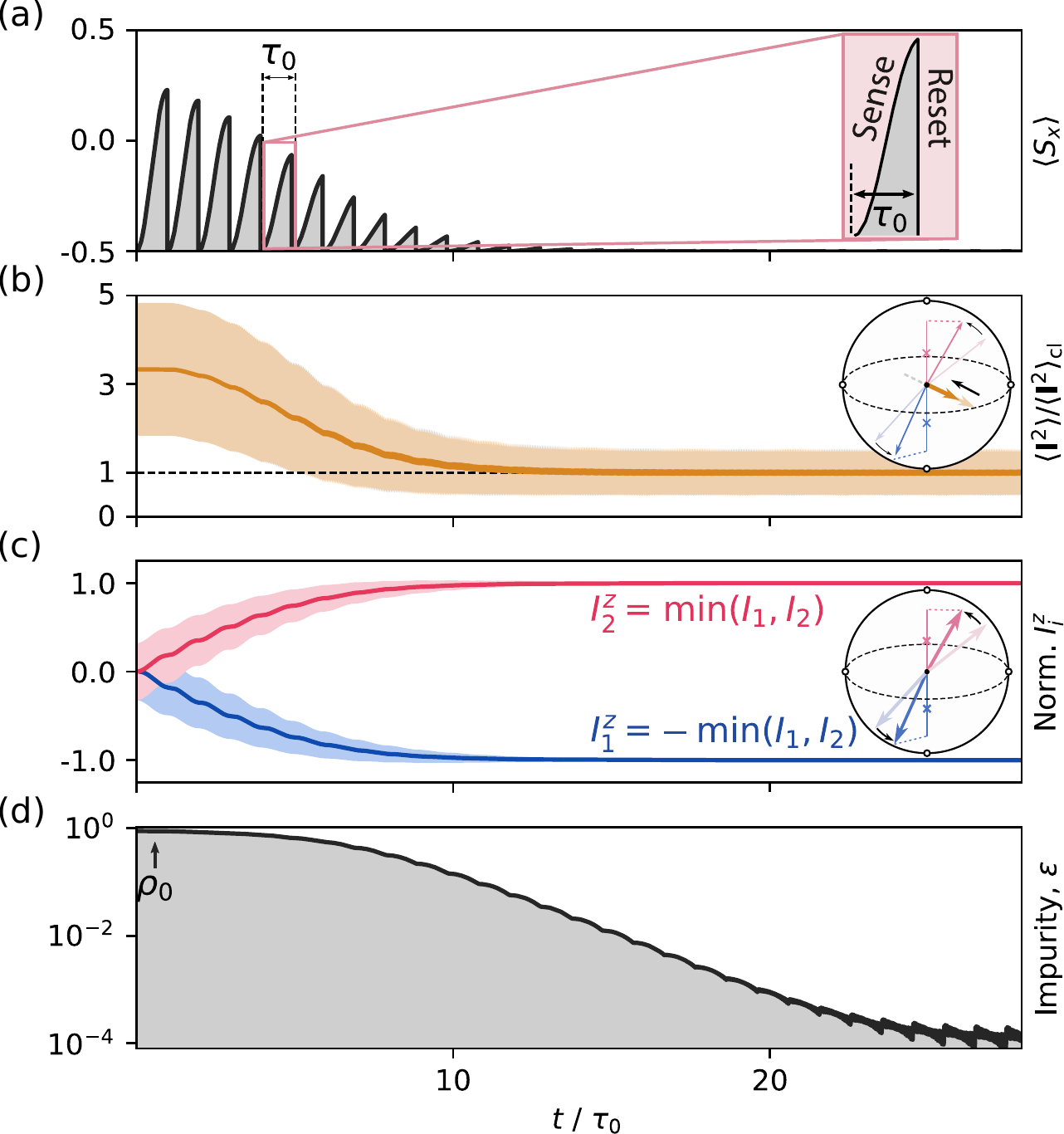}
    \caption{\textbf{Ideal system's dynamics for $N=32$, and $I_1=I_2=\sqrt{32/2}$.} \textbf{a,} The expectation value of $S_x$ as the function of the protocol time.   \textbf{Inset}, $\langle S_x \rangle$ during a single iteration of sensing and instantaneous reset (c.f. the middle and the right-most panels of the Fig.~\ref{fig:my_label}e).
    \textbf{b,} Magnitude (solid orange line) and the uncertainty (shaded area) of $\langle \mathbf{I}^2 \rangle$ as a function of the protocol time, normalized by $\langle \mathbf{I}^2 \rangle_{\text{cl}}=(I_1+I_2)(|I_1-I_2|+1)$. \textbf{Inset}, Suppression of the total $I$ (orange arrow) down to the classical limit. 
    \textbf{c,} The normalized expectation values of $I_1^z$ and $I_2^z$ (solid blue and red lines, respectively) and their uncertainties (corresponding shaded areas) as the functions of the protocol time. The normalization involves dividing the $y$-axis values by $\min(I_1,I_2)$. \textbf{Inset}, $\mathbf{I}_1$ and $\mathbf{I}_2$ dynamics towards a pure anti-polarized state. 
    \textbf{d,} The bath impurity, $\epsilon$, as a function of the protocol time. 
    $\rho_0$ denotes the state of the bath at the beginning of the second stage of the protocol. The $x$-axis is shared across the panels. 
   } 
    \label{fig:my_label2}
\end{figure}

\subsection{Ideal system dynamics}
We verify the convergence of the second stage of the protocol towards the ground state by calculating the full quantum evolution of the system numerically. We treat the ideal case, for which the three-body interaction is fully coherent and the central spin reset is instantaneous. We repeat this second stage as many times as necessary to reach steady state. For convergence towards the steady state under reasonable computational resources, we choose to work with $N=32$ bath spins and focus on the $I_1=I_2=\sqrt{32/2}$ manifold \cite{supplementary}. We take the first stage of the protocol to be fully capable of confining the ensemble's dynamics to the $I^z=0$ subspace \cite{Jackson2022}, spanned by $\{\ket{I_1^z=-M+n,I_2^z =M-n}, \quad n=0,1,2,..,2M\}$, to which we accordingly restrict our simulation. At the beginning of the simulation, we set the ensemble's density operator, $\rho$, to an equal statistical mixture of the subspace basis states. 

Figure~\ref{fig:my_label2}a shows the central spin's evolution under the activated three-body interaction and confirms the coherent oscillation in the $\{\ket{\downarrow_x},\ket{\uparrow_x}\}$-basis, with a half-period of $\tau_0\sim \tfrac{2\pi\omega_{\text{c}}}{a^2 (I_1 \times I_2)}$ -- inversely proportional to the fastest three-body interaction rate \cite{supplementary}. The state of the central spin is instantaneously reset every $\tau_0$, and during the consecutive iterations the magnitude of the acquired $\ket{\uparrow_x}$-population drops down to zero as the ensemble approaches its ground state. Semi-classically, this corresponds to a drop in the magnitude of the noise $\langle \mathbf{I}^2 \rangle$ sensed by the central spin (Fig.~\ref{fig:my_label2}a). We verify this independently as shown in Fig.~\ref{fig:my_label2}b, where $\langle \mathbf{I}^2 \rangle$ saturates to its classical ${\sim }\sqrt{N}$ limit as steady state is reached. Figure~\ref{fig:my_label2}c shows the polarizations of each of the two species, $I_1^z$ and $I_2^z$ as a function of iteration time. We see that they saturate to maximal and opposite values $I_1^z = +\min(I_1,I_2)$ and $I_2^z = -\min(I_1,I_2)$. Importantly, at the longest iteration times their uncertainties approach zero, as expected for a pure anti-polarized state.

We quantify the quality of state preparation following our protocol using the bath state impurity:
\begin{equation}
    \epsilon = 1 - \Tr \rho^2,
\end{equation}
where $\rho$ is the density operator of the bath. For a pure state this measure is known to reach zero. In our simulation, we selected a directionality parameter of $\kappa=5$ (Eq.~\ref{eqn:kappa}) as an example case. Figure~\ref{fig:my_label2}d shows that for this value of $\kappa$ the impurity reaches a steady-state value of $\epsilon\approx10^{-4}$, which represents a negligible initialization error. After settling, the system is trapped in an oscillatory limit-cycle, resulting from the competition of $\ket{\downarrow_x}\ket{I_1^z, I_2^z}\leftrightarrow \ket{\uparrow_x}\ket{I_1^z - 1, I_2^z + 1}$ and $\ket{\downarrow_x}\ket{I_1^z, I_2^z}\leftrightarrow\ket{\uparrow_x}\ket{I_1^z + 1, I_2^z - 1}$ transitions. This error $\epsilon$ can be made arbitrarily small by increasing $\kappa$, which we address in the following section.

\subsection{Dependence of protocol performance on system parameters}

In the absence of dephasing, there remains a fundamental trade-off between impurity $\epsilon$ and the convergence time to steady-state $T_{\mathrm{c}}$. This is because reducing the three-body interaction -- increasing $\kappa$ -- ensures a smaller contribution from off-resonant processes, thus reducing the initialization error, but slows the dynamics to steady-state. For $\kappa \to \infty$ one could expect bringing the impurity $\epsilon$ arbitrarily close to zero, but this comes with a prohibitively long convergence time, $T_{\mathrm{c}}$. 

\begin{figure}
    \centering
    \includegraphics[width=\columnwidth]{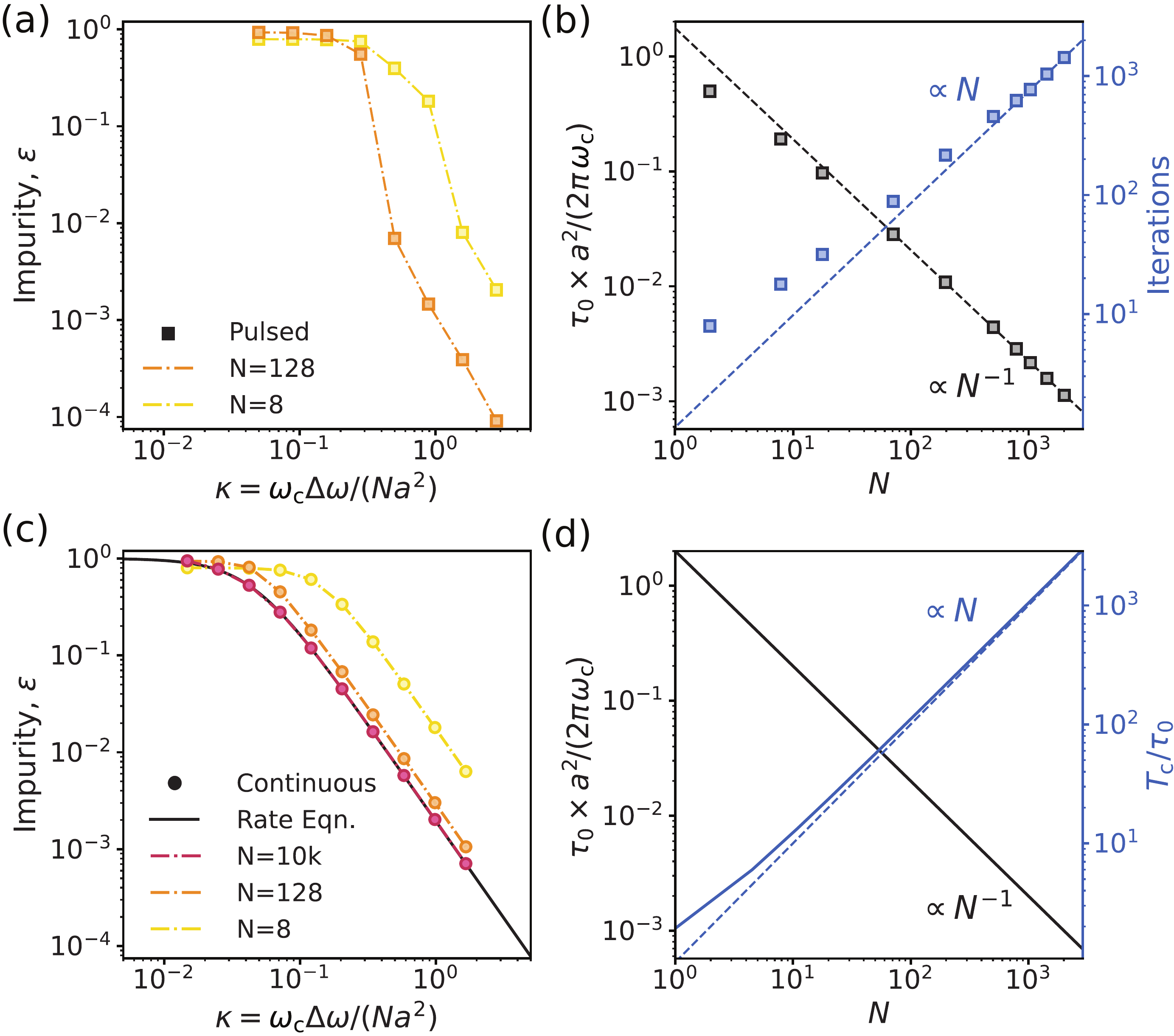}
    \caption{\textbf{Dependence on system parameters.} \textbf{a,} Steady state impurity, $\epsilon$, as a function of the directionality parameter, $\kappa=\omega_{\text{c}}\Delta \omega  / (Na^2 )$. 
    \textbf{b,} Normalized three-body interaction half-periods (i.e. the pulse durations), $\tau_0$ (black squares), and the numbers of stage iterations (blue squares) required to reach convergence in the pulsed protocol, as a function of $N$.  The number of iterations was calculated as three times the $1/e$-drop time in the impurity's exponent to the settled value. The dashed lines illustrate the asymptotic behavior, $N\to\infty$, of both quantities.
    \textbf{c,} The $\kappa$ dependence of impurity, $\epsilon$, in the continuous protocol. 
    The solid black line displays a corresponding solution from the rate equation model, which correctly captures the high-$N$ limit.
    \textbf{d,} Prediction of the convergence results in the Fig.~\ref{fig:my_label3}b using the rate equation model.}
    \label{fig:my_label3}
\end{figure}

To visualize the achievable steady-state impurities, and related convergence times, we run a complete simulation of our pulsed protocol for a range of values of $\kappa$ and for $N=8$ and $N=128$. Figure~\ref{fig:my_label3}a confirms the clear trend of purity improvement with an increase in $\kappa$, where substantial degrees of purification are achieved past $\kappa \gtrsim 1$. Figure~\ref{fig:my_label3}b shows the inverse $N$-dependence of the half-period of the collective three-body exchange, $\tau_0 \sim 4\pi \omega_\text{c}/Na^2$, and the number of stage iterations required to reach convergence proportional to $N$. The convergence time is a simple product of these two quantities and thus combines to 
\begin{equation}
T_{\mathrm{c}} \sim 16 \pi \omega_{\text{c}}/a^2,    
\end{equation}
indicating that the convergence time is dependent purely on the three-body interaction strength, and not the system size. 

Verifying the above trends in the large-$N$ limit becomes computationally prohibitive. As an efficient way to extend our results into this regime, we turn to a steady-state solver of a quantum master equation in which the coherent three-body exchange proceeds continuously and simultaneously with a central spin reset whose rate is $\Gamma_{\text{op}}=2 \pi/ \tau_0$. Figure~\ref{fig:my_label3}c displays the steady state impurity, $\epsilon$, for $N$ up to $10,000$ and for the same range of values of $\kappa$ as in Fig.~\ref{fig:my_label3}a. This confirms effective purification for $\kappa \gtrsim 1$ for large $N$. The solid black curve in Fig.~\ref{fig:my_label3}c is the prediction from a simple rate equation\cite{supplementary}, i.e. $N\rightarrow\infty$, which overlaps with the steady-state model in the large-$N$ limit. Both models feature a $\epsilon \sim\kappa^{-2}$ roll-off, as obtained analytically from a first-order expansion of the impurity in the rate equation model (see \cite{supplementary}). As shown in Fig.~\ref{fig:my_label3}d, eigenvalue analysis of the rate equation allows us to calculate the convergence time, $T_{\mathrm{c}}$, agreeing with the size-independent $T_{\mathrm{c}} \sim 16 \pi \omega_{\text{c}}/a^2$ behavior observed in the pulsed model (Fig.~\ref{fig:my_label3}b). 

Comparing the pulsed (Fig.~\ref{fig:my_label3}a) and continuous (Fig.~\ref{fig:my_label3}c) protocol performance, we note that maintaining the temporal separation between the central spin reset and activation of the three-body interaction with the ensemble allows reaching lower steady-state impurities. This property, illustrated by comparing experiments done in the continuous \cite{Gangloff2019} and pulsed \cite{Jackson2022} regimes, can be straight-forwardly explained: continuously measuring the central spin reduces its ability to sense the ensemble noise, and in turn limits the achievable purity.

\subsection{Resilience to system imperfections}

\begin{figure}
    \centering
    \includegraphics[width=\columnwidth]{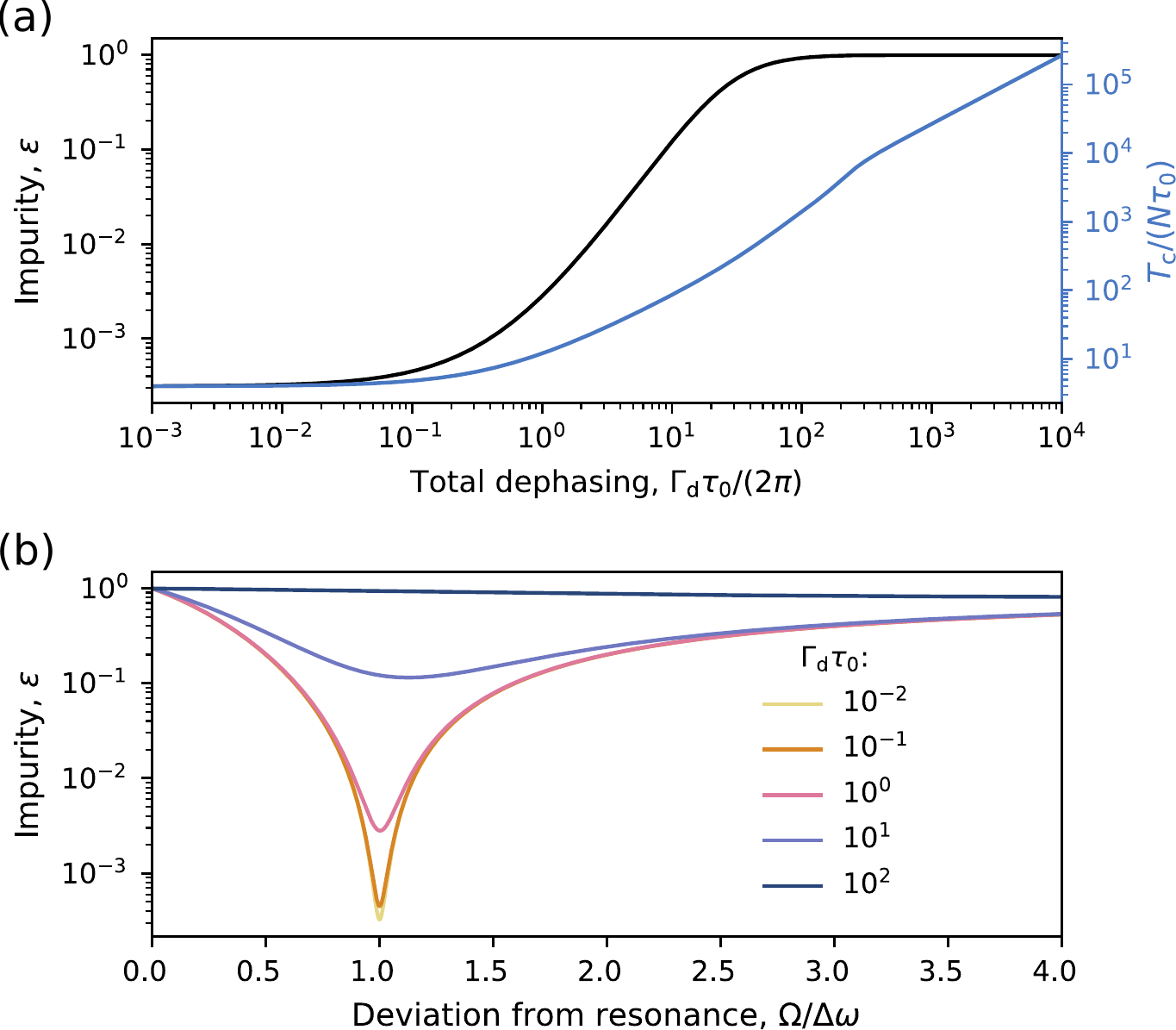}
    \caption{\textbf{Resilience to dephasing and imperfections.}  
    \textbf{a,} Impurity (black curve) and convergence time (blue curve) of the protocol as a function of the normalized total dephasing rate, $\Gamma_{\mathrm{d}}\tau_0/(2\pi)$, calculated at the $\Omega = \Delta \omega$ resonance. 
    \textbf{b,} Impurity of the steady state as a function of the deviation from the $\Omega = \Delta \omega$ resonance, in the presence of an increasing amount of dephasing. $\kappa=10$ for all the curves in the plots.  
    } 
    \label{fig:my_label4}
\end{figure}

Real physical systems deviate from the idealized behavior considered so far, as they are affected by central spin and ensemble inhomogeneous dephasing, as can occur if the ensemble spins have a spread of Larmor precession frequencies $\sqrt{\Delta^2 \omega_i}>0$ \cite{Jackson2021,Jackson2022,supplementary}. Each stage of our purification protocol relies on the exchange dynamics between the central spin and the ensemble. Working in this strong coupling regime renders the protocol equally sensitive to both the central spin and ensemble dephasing mechanisms\cite{supplementary}, proceeding at rates $\Gamma_{\mathrm{c}}$ and $\Gamma_{\mathrm{b}}$, respectively. We thus use the total dephasing rate $\Gamma_{\mathrm{d}} = \Gamma_{\mathrm{c}} + \Gamma_{\mathrm{b}}$ as the relevant parameter and explore the robustness of our protocol against such system imperfection. 

Using the rate equation approach \cite{supplementary}, we calculate the steady-state impurity, $\epsilon$, as a function of the dephasing rate normalized by the three-body interaction rate, $\Gamma_{\mathrm{d}}\tau_0/(2\pi)$, as shown in Fig.~\ref{fig:my_label4}a (black curve). A significant alteration of the protocol performance is observed when the three-body interaction time $\tau_0$ becomes longer than the typical dephasing time, i.e. $\Gamma_{\mathrm{d}} \tau_0/(2\pi) \gtrsim 1$. Nonetheless, for large values of the directionality parameter $\kappa$ (here, $\kappa=10$), the engineered steady state impurity remains lower than unity by orders of magnitude. In this low-impurity regime (and for $N\to \infty $), this is quantitatively captured by a simple expression \cite{supplementary}:
\begin{equation}
\epsilon \approx \frac{2}{1+64\kappa^2[1+2\Gamma_{\mathrm{d}}\tau_0/(2\pi)]^{-2}}.
\end{equation}  
The rate of the activated three-body exchange is also affected by dephasing, which results in an increased protocol convergence time, $T_{\mathrm{c}}$ (blue line in Fig.~\ref{fig:my_label4}a). As expected from a time-energy uncertainty principle, dephasing also comes with a broadening of the three-body resonance $\Omega = \Delta \omega$, as seen in Fig.~\ref{fig:my_label4}b.

\subsection{Benchmarking candidate material platforms}
\begin{figure}[t!]
    \centering
    \includegraphics[width=\columnwidth]{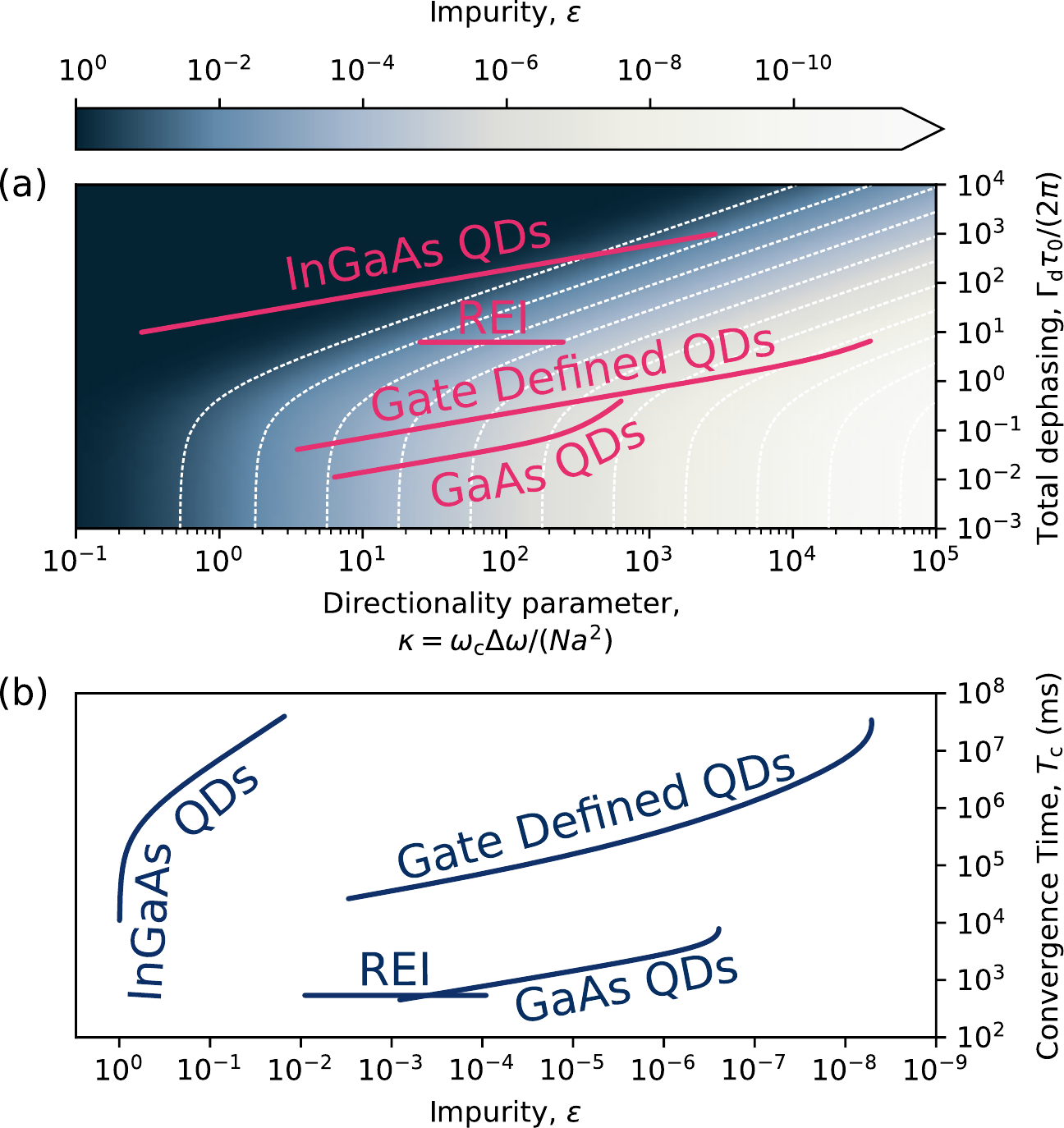}
    \caption{\textbf{Benchmarking candidate material platforms} \textbf{a}, Impurity as a function of the platform-agnostic parameters, $\kappa$ and the normalized total dephasing $\Gamma_\mathrm{d}\tau_0/(2\pi)$. White dashed lines show the contours of constant impurities increasing exponentially from $10^{-1}$ to $10^{-11}$ by a factor $10$. Solid violet lines correspond to the typical operating regimes for selected candidate platforms. \textbf{b}, Summary of the achievable steady-state impurities and convergence times for the candidate platforms.} 
    \label{fig:my_label5}
\end{figure}
Throughout this work, we have identified the two platform-agnostic control parameters determining the protocol performance: the directionality parameter, $\kappa$, and the normalized total dephasing rate, $\Gamma_\mathrm{d}\tau_0/(2\pi)$. We now evaluate these parameters for candidate physical systems and quantify the corresponding bounds on steady-state bath impurities, $\epsilon$, as well as the protocol convergence times, $T_{\mathrm{c}}$, as shown in Fig.~\ref{fig:my_label5}. 

We restrict our analysis to the physical platforms that naturally realize dense central spin systems (see the Hamiltonian of Eq.~\ref{eq:Hamiltonian_CSS}), and in which the rudimentary protocol ingredients, like the control and reset of the central spin state, have been demonstrated previously. Our non-exhaustive selection of the candidate systems includes Gate-Defined Quantum Dots (QDs), Lattice matched GaAs-AlGaAs QDs, Stranski-Krastanow InGaAs QDs, and Rare Earth Ions (REI). The system-specific parameters used in the calculations are tabulated in the supplementary materials \cite{supplementary}.

For QD systems the two spin species that break ensemble symmetry are simply two nuclear-spin species with different gyromagnetic ratios: gallium and arsenic. Calculations for QD systems have taken into account the spin-orbit magnetic-field $B$ dependence of the central-spin lifetime, $\propto B^{3}$\cite{PhysRevLett.100.046803} and $\propto B^{5}$\cite{PhysRevB.81.035332} for the Gate Defined QDs and the epitaxial QDs (both InGaAs and GaAs-AlGaAs), respectively. Beyond magnetic fields of $10$ T, this spin-orbit effect becomes the most performance-limiting factor and caps the impurity-convergence time trade-off.

Spectacular purification to $\epsilon < 10^{-4}$ can be achieved for GaAs and Gate Defined QDs, due to the high values of the directionality parameter, $\kappa$, and the low total dephasing rates, $\Gamma_{\mathrm{d}}\tau_0/(2\pi)$, within the typical range of externally applied magnetic fields, $B$ (see Fig.~\ref{fig:my_label5}a). Gate-Defined QDs feature ${\sim}100$-fold larger directionality parameters, $\kappa$, than the GaAs QDs at the same externally applied magnetic field, for the same (unnormalized) total dephasing rates, $\Gamma_{\mathrm{d}}$ \cite{Bluhm2011,Zaporski2022}. This generally leads to higher degrees of purification, but significantly longer convergence times for Gate Defined QDs (see Fig.~\ref{fig:my_label5}b). 

InGaAs QDs feature a larger effective nuclear dephasing rate, $\Gamma_{\mathrm{d}}$, due to nuclear spectral inhomogeneity arising from strain-induced quadrupolar broadening \cite{Stockill2016}. In their case, three-body interactions could still lead to a weak purification, as consistent with the observation of enhanced spin-wave modes at low ensemble polarization \cite{Gangloff2021}.

We now turn our attention to REI systems - specifically to ${}^{171}\text{Yb}^{3+}:\text{YVO}_4$, used recently to demonstrate quantum-state transfer between a ${}^{171}\text{Yb}$ electronic central spin and the second shell of the nearest ${}^{51}V$ nuclei\cite{Ruskuc2022}.
This system belongs to the regime of small but dense central-spin systems and offers little tunability with an external magnetic field as all nuclear spins are of the same species. However, the nuclear-spin shells surrounding the central spin can be distinguished via quadrupolar shifts, allowing a set of values for the effective $\Delta \omega$\cite{supplementary}. 
The required central-spin mediated three-body interaction arises as a second-order magnetic dipole-dipole coupling between two of the first, second, and higher shells of ${}^{51}V$ nuclei interfaced with the
${}^{171}\text{Yb}$. The measured nuclear coherence time is three times shorter than the interaction time $\tau_0$, but this still allows for generating high-purity anti-polarized states (see Fig.~\ref{fig:my_label5}).

\subsection{Preparation of a many-body singlet}

\begin{figure*}
    \centering
    \includegraphics[width=2\columnwidth]{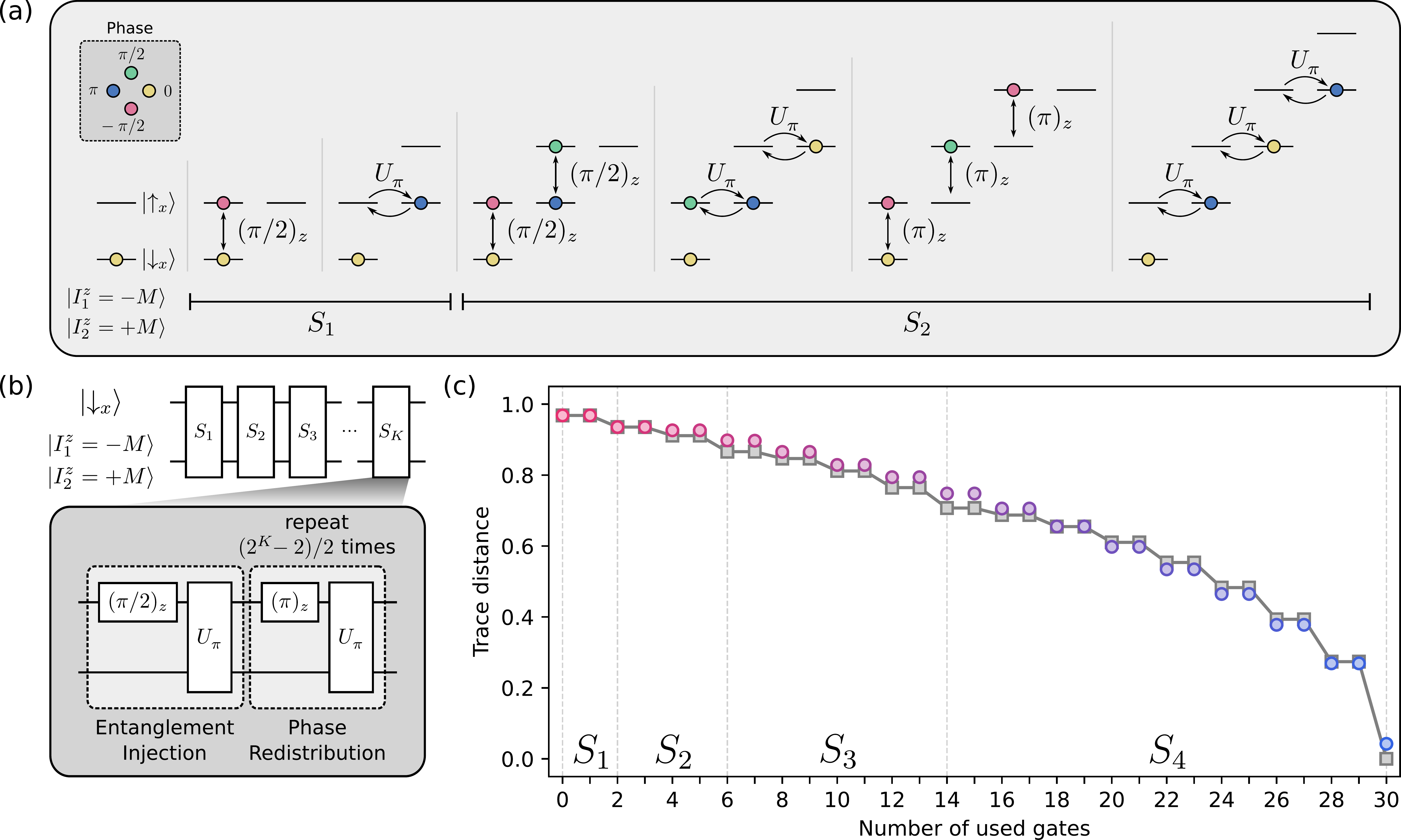}
    \caption{
    \textbf{Reaching a quantum limit} \textbf{a,} Action of the unitary gates (black arrows) along the effective Jaynes-Cummings ladder of $\ket{I_1^z=-M+n,I_2^z=M-n}$-states, where $M=\min(I_1,I_2)$. The illustrated sequence of six gates prepares a many-body singlet for $K=2$. The inset displays a relative phase color-coding applied throughout the panel. The global phase is factored out, leaving the lowest energy state with a zero reference phase. 
    \textbf{b,} Concatenation of the composite $S_j$ gates turns the anti-polarized state into a many-body singlet. The gray box contains a quantum circuit corresponding to the $S_{K}$ gate, for an arbitrary integer $K$.
    \textbf{c,} Trace distance from a singlet state as a function of the number of singlet-preparing gates used within the simplified (gray squares) and the real (circles) systems with $K=4$ (i.e. $N=112$), following the exact and variational-searched protocols, respectively. The red-to-blue gradient indicates that the structure of the optimal protocol (discussed in the Ref. \cite{supplementary}) varies with the number of used gates. 
   } 
    \label{fig:my_label6}
\end{figure*}

The singlet state $\ket{I=0}$ is a superposition of all $\ket{I_1^z=-M+n,I_2^z=M-n}$ eigenstates with a $\propto(-1)^n$ phase on each state, as shown in Eq.~\ref{eq:singlet-super position}. Having established a pure anti-polarized state of the two sub-ensembles (i.e. $n=0$) following the first two stages of the protocol, the third and final stage will prepare the singlet state by weaving an alternating phase into the Jaynes-Cummings ladder, as shown in Fig.~\ref{fig:my_label6}a. We stress that the many-body singlet state is \emph{not} an eigenstate of our system Hamiltonian; however, it refocuses every $\Delta t=2\pi/\Delta \omega$ following the protocol termination\cite{supplementary}.

We consider only the ideal execution of this final stage involving unitary gates, free of any dephasing. We work in a frame co-rotating with $\Omega S_x + \Delta \omega(I_1^z-I_2^z)/2$. For the sake of clarity, we outline the ideal protocol steps assuming that the effective Jaynes-Cummings ladder has $2M+1=2^K$ rungs, corresponding to a total number of ensemble spins $N\sim(2^K-1)^2/2$; the generalization to an arbitrary $N$ is straightforward. In the first instance, we take a simplified scenario in which the three-body-interaction is independent of the $\ket{I_1^z=-M+n,I_2^z=M-n}$-state; we will later take the dependence on $n$ into account.   

We apply the following sequence of unitary operations, starting from the state

\begin{equation}
\ket{\psi_0} = \ket{\downarrow_x}\ket{I_1^z=-M,I_2^z=M}\text{,}
\end{equation}
\noindent
(i) a central spin $(\pi/2)_z$-gate, giving state

\begin{equation}
\frac{1}{\sqrt{2}}\left(\ket{\downarrow_x}-i\ket{\uparrow_x}\right)\ket{I_1^z=-M,I_2^z=M}\text{,}
\end{equation}
as illustrated in the second level diagram from the left in Fig.~\ref{fig:my_label6}a;

(ii) a three-body-interaction $\pi$-gate, $U(\tau)=U_\pi$, where $\tau$ is the interaction time, and which up to a global phase leaves the system in the state

\begin{equation}
\begin{split}
\ket{\psi_1} = \frac{1}{\sqrt{2}}\ket{\downarrow_x}&(\ket{I_1^z=-M,I_2^z=M}-\\ &\ket{I_1^z=-M+1,I_2^z=M-1}) \text{,}
\end{split}
\end{equation}
as shown in the third level diagram from the left in Fig.~\ref{fig:my_label6}a.
\noindent
Together, this pair of gates $(\pi/2)_z$ and $U_\pi$ injects entanglement into the spin ensemble via the central spin. We combine them into a composite gate $S_1$, as shown in Fig.~\ref{fig:my_label6}a. Application of this gate doubles the overlap with the singlet state, $|\bra{\psi_1}\left(\ket{\downarrow_x}\ket{I=0}\right)|^2$, from the initial state's, $|\bra{\psi_0}\left(\ket{\downarrow_x}\ket{I=0}\right)|^2$. To increase this overlap further, we apply an extended composite gate, $S_2$, which contains two steps: (i) entanglement injection using $S_1$ (the fourth and fifth level diagrams from the left in Fig.~\ref{fig:my_label6}a) and (ii) phase redistribution onto the 
$\ket{I_1^z=-M+n,I_2^z=M-n}$-states with $n=0,1,2,3$ using a central spin $(\pi)_z$-gate and a three-body-interaction $\pi$-gate (the two right-most level diagrams in Fig.~\ref{fig:my_label6}a). We generalize this sequence to a composite gate $S_K$, for which the phase redistribution is applied $2^{K-1}-1$ times (gray box in Fig.~\ref{fig:my_label6}b). Each application of a phase redistribution sub-sequence brings the next highest rung on the ladder into the ensemble superposition state. As a result, the overlap of system state $\ket{\psi_j}$ with the many-body singlet state $|\bra{\psi_j}\left(\ket{\downarrow_x}\ket{I=0}\right)|^2$ doubles at each step of a sequence of composite gates $S_j, \text{ for } j=1,2,..,K$. 

The modular structure of this algorithm is advantageous in terms of minimizing the impact of the central spin dephasing on the ensemble state preparation. It is sufficient for the central spin to stay coherent over a given $S_j$ gate duration, after which it can be safely reinitialized. 
We note that the phase redistribution is the most operation-costly sub-sequence, as it involves $2^j-2$ gates within each $S_j$ gate. Nevertheless, the protocol complexity remains linear with the number of states along the ladder $2^K$ as the total number of gates in a complete sequence is $2(2^K-1)$.

We demonstrate the performance of this algorithm in Fig.~\ref{fig:my_label6}c, using the trace distance from the engineered state to the singlet state as a function of the total number of gates (gray squares). For $K=4$ ($N=112$), we can visualize the steady progression from $\ket{\psi_0}$ towards the singlet state $\ket{\downarrow_x}\ket{I=0}$ as the algorithm steps through the sequence of composite gates $S_1$ to $S_4$, culminating in an exact preparation where the final trace distance reaches zero. The classical limit $\langle\mathbf{I}^2\rangle_{\text{cl}} \sim \sqrt{N}$ is overcome as soon as the sequence begins to inject entanglement into the ensemble, and reaches the quantum limit of $\langle\mathbf{I}^2\rangle_{\text{qu}}=0$ at its termination. Interestingly, this happens at the expense of raising the uncertainty in $I_1^z$ and $I_2^z$ towards their thermal values ${\sim} \sqrt{N}$, akin to squeezing.

We now turn to the more realistic description of the spin ensemble for which the three-body-interaction $\pi$-gate time depends on the state $\ket{I_1^z=-M+n,I_2^z=M-n}$. This interaction time will take on values from ${\sim}2M$ (for $n=0,2M$) to ${\sim} M(M+1)$ (for $n=M$) across the ladder\cite{supplementary}. While this behavior complicates the algorithm implementation, the core structure used to inject entanglement and redistribute phase remains in place. With the right choice of the three-body interaction times $\tau$ and the central spin $z$-gate phases, $\phi$, it is possible to reach the singlet state. To do so we treat the sequence parameters $\{\tau_i,\phi_i\}$ variationally for each gate to minimize the (Hilbert-Schmidt) trace distance to the singlet state from

\begin{equation}\label{ansatz}
\begin{split}
    \ket{\psi}= \prod^{\xleftarrow[]{}}_{i=0} \Big(&U(\tau_i)\cdot(\phi_i)_z \Big)\\ &\ket{\downarrow_x} \ket{I_1^z=-M,I_2^z=M}.
\end{split}
\end{equation}

Formally, this involves an optimization over a $\sim 2^K$-dimensional space of parameters, for which we employ a gradient-descent algorithm \cite{supplementary,KHANEJA2005296}. The resulting trace distances for the sequences of increasing length are presented in Fig.~\ref{fig:my_label6}c as the colored circles (red-to-blue gradient is consistent with the color-coding used in ref. \cite{supplementary}). Strikingly, our algorithm arrives within a trace distance of a few $\%$ from the singlet state. A process optimization in larger systems, aided by machine learning algorithms, could prove to be a viable route for creating arbitrary state superpositions\cite{Cerezo2021}.

\section{Conclusions and Outlooks}

In this work, we have proposed a protocol that can initialize a spin ensemble into a pure anti-polarized state and steer it towards a many-body entangled singlet state -- exclusively by using a single central-spin qubit. To do so, we have made use of the three-body interaction naturally present in dense spin ensembles and shown that it can be harnessed by breaking the ensemble into two spin species. We have suggested several platforms where this algorithm would be realizable, and where significant purity can be achieved even in the presence of dephasing. We note that in these systems, breaking the spin ensemble can take multiple forms including, but not limited to, nuclear-spin species with different gyromagnetic ratios \cite{Stockill2016} or high-spin species which split into two effective qubit ensembles under the influence of electric-field gradients (e.g. from strain) \cite{Chekhovich2020}.

From the perspective of an electron spin qubit hosted in a material with non-zero nuclear spins, a singlet state of its surrounding spin ensemble would dramatically boost its coherence -- both homogeneous and inhomogeneous noise sources\cite{Stockill2016} would be quenched altogether. From the perspective of leveraging this spin ensemble as a quantum memory resource\cite{Taylor2003}, initialization to a pure collective state is sufficient to run an algorithm with unit fidelity, and the availability of two anti-polarized species could even be extended to a two-mode register. The state-engineering recipes that we have established could be extended to more elaborate computational\cite{Anikeeva2020} and error-correcting \cite{Abobeih2022} algorithms. Fundamentally, tracking a many-body state in the presence of tuneable interactions can reveal the entanglement dynamics in and out of the central-spin system, opening an experimental window onto quantum information scrambling and area laws for entanglement entropy.

\textbf{Acknowledgements:} We acknowledge support from the US Office of Naval Research Global (N62909-19-1-2115) and the EU H2020 FET-Open project QLUSTER (862035). D.A.G. acknowledges a Royal Society University Research Fellowship, and C.LG. a Dorothy Hodgkin Royal Society Fellowship. L.Z. acknowledges support from the EPSRC DTP.  We also thank S. Economou and E. Barnes for fruitful discussions.

\bibliography{singlet}

\end{document}


	
\title{Supplementary Materials for: A many-body singlet prepared by a central spin qubit
}
\author{Leon Zaporski\textsuperscript{1,*}}
\author{Stijn R. de Wit\textsuperscript{1,2,*}}
\author{Takuya Isogawa\textsuperscript{1}}
\author{Martin Hayhurst Appel\textsuperscript{1}}
\author{Claire Le Gall\textsuperscript{1}}
\author{Mete Atat\"ure\textsuperscript{1,$\dagger$}}
\author{Dorian A. Gangloff\textsuperscript{3,$\dagger$}}

\noaffiliation

\affiliation{Cavendish Laboratory, University of Cambridge, JJ Thomson Avenue, Cambridge, CB3 0HE, UK}
\affiliation{MESA+ Institute for Nanotechnology, University of Twente, The Netherlands}
\affiliation{Department of Engineering Science, University of Oxford, Parks Road, Oxford, OX1 3PJ
\\ \ \\
\textsuperscript{*}\,These authors contributed equally\\
\textsuperscript{$\dagger$}\,Correspondence to: ma424@cam.ac.uk, dorian.gangloff@eng.ox.ac.uk.
\\ \ \\
}
	
	\maketitle
	\tableofcontents

\section{Manifold degeneracies}

\begin{figure}
\includegraphics[width=0.5\textwidth]{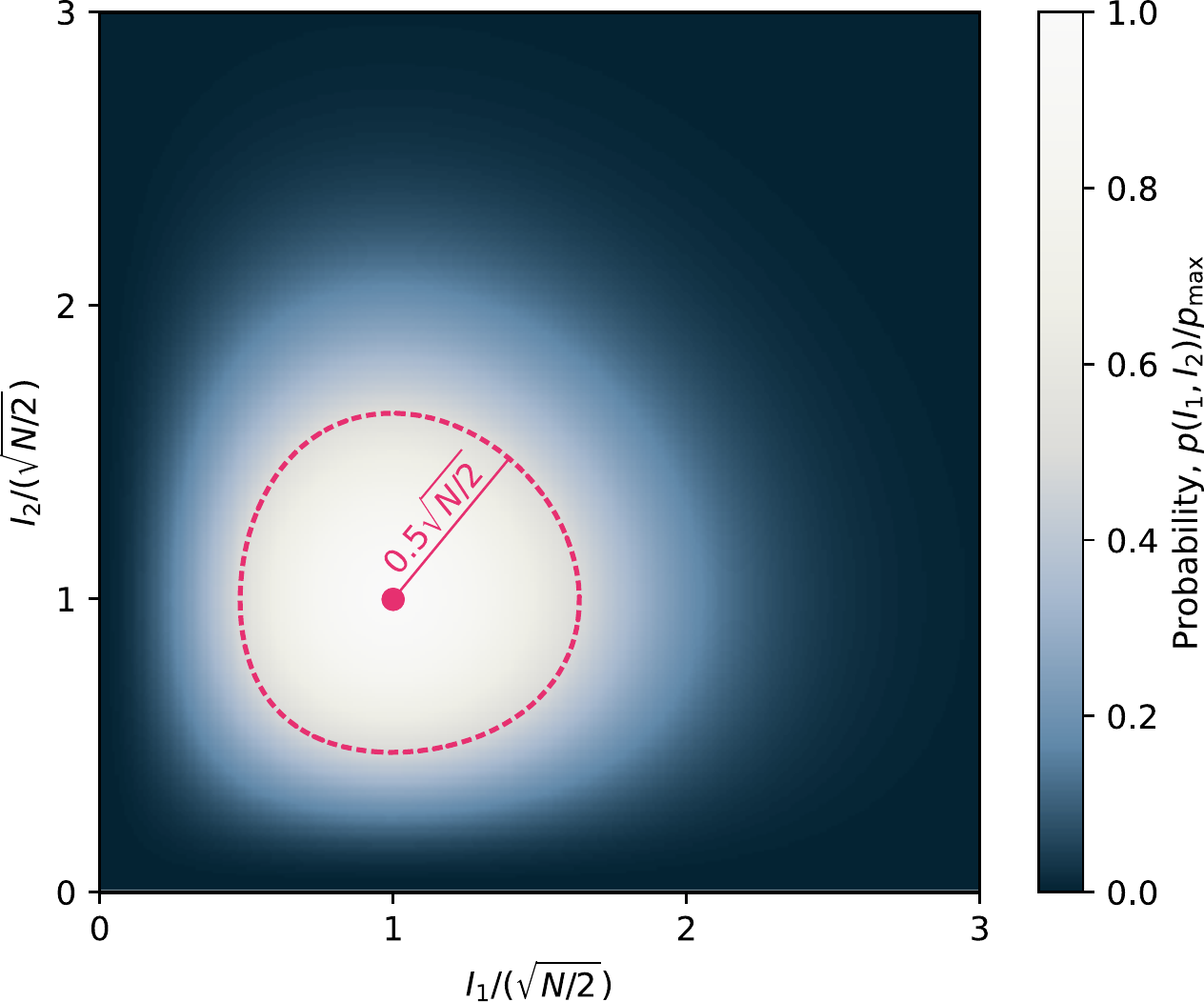}
		\caption{\textbf{Probability distribution of sampling an $I_1,I_2$-manifold in an infinite-temperature ensemble.
} The pink dashed curve represent a half-maximum contour of $\sqrt{N/2}$ diameter. The pink point represents the mode of the distribution at $I_1=I_2=\sqrt{N/2}$.} 
	\label{fig:Degeneracies}
\end{figure}
 
Within the manuscript we make a statement about the dynamics in $I_1=I_2=\sqrt{N/2}$ manifold being representative of the complete ensemble dynamics. This follows from the $(I_1,I_2)$-manifold degeneracy, captured by the:
\begin{equation}
p(I_1,I_2)\propto I_1(I_1+1) \times I_2(I_2+1) \times e^{-2(I_1^2+I_2^2)/N}
\end{equation}
probability distribution, for an ensemble initially at infinite temperature\cite{Jackson2022}. As shown in Fig.\ref{fig:Degeneracies}, this distribution features a peak at $\sqrt{N/2}$ of width $\propto \sqrt{N/2}$, and an exponentially-suppressed high-$I_i$ tail. 
Therefore, when randomly sampling the $I_1,I_2$-distribution in the experiment, a majority of the experimental runs, will satisfy $I_1,I_2 \sim \sqrt{N/2}$. 

A typical deviation from $I_1=I_2=\sqrt{N/2}$ features $I_1 \ne I_2$. In the sec. \ref{sec:fullquantumS}, we study the protocol performance in the $I_1\ne I_2$ case, and compare it to the $I_1=I_2$ case (studied in the manuscript) to find the same degree of state purification, and identical behaviour of the central spin throughout the protocol.  


However, the robustness of the optimized unitary gate sequence in the third stage of the protocol is expected to be more sensitive on sampled values of $I_1$ and $I_2$. This effect could be mitigated following the generalization of our variational optimization procedure to a larger Hilbert space. In a proof-of-concept experiment, it could be also bypassed with a measurement post-selection or measurement-based feed-forward, provided the possibility of a single-shot read-out of the central spin state. 
For example, measuring the transition rate asymmetry\cite{Gangloff2021} of the $\Omega=\omega_1$ activated processes:
\begin{equation}
\begin{split}
&\ket{\downarrow_x}\ket{I^z_1,I^z_2}\to \ket{\uparrow_x}\ket{I^z_1-1,I^z_2}\\
&\ket{\uparrow_x}\ket{I^z_1,I^z_2}\to \ket{\downarrow_x}\ket{I^z_1+1,I^z_2}
\end{split}
\end{equation}
and the transition rate asymmetry of the  $\Omega=\omega_2$ activated processes:  
\begin{equation}
\begin{split}
&\ket{\downarrow_x}\ket{I^z_1,I^z_2}\to \ket{\uparrow_x}\ket{I^z_1,I^z_2-1}\\  &\ket{\uparrow_x}\ket{I^z_1,I^z_2}\to \ket{\downarrow_x}\ket{I^z_1,I^z_2+1} 
\end{split}
\end{equation}
constrains both $I_1$ and $I_2$, which can be used to conditionally discard the post-protocol state, or to adjust the gate parameters in the third stage of the protocol. 


	\section{Full simulation of quantum dynamics}
    To simulate the quantum dynamics of the composite system in a reduced Hilbert space (see the manuscript) we use a master equation and steady state solvers from the QuTiP\cite{JOHANSSON20121760,JOHANSSON20131234} library in Python. 

    The exchange between the central spin and the ensemble is modelled by the following Hamiltonian, written in a frame rotating with a $\delta$-detuned laser drive, following a Rotating Wave Approximation:
    \begin{equation}
    \begin{split}
    H(t)=&\Omega(t) S_x +\delta S_z + \sum_{i=1,2} \Big(\omega_i -\frac{a^2}{4\omega_{\text{c}}}\Big)I_i^z + \sum_{i=1,2}aS_zI_i^z\\ &+ \sum_{i,j=1,2} \frac{a^2}{4\omega_{\text{c}}} S_z(I_i^+ I_j^-+I_i^- I_j^+) + \sum_{i=1,2} \frac{a\nu_i}{2\omega_i} S_z I_i^x.
    \end{split}
    \end{equation}
    The individual terms are defined in the manuscript. 

    Within the first stage of the protocol setting $\Omega(t)=\tfrac{1}{2}(\omega_1+\omega_2)$ activates the non-collinear ($\propto S_z I_i^x$) interaction. Within the second and third stages of the protocol, $\Omega(t)=\Delta \omega$ activates the three-body interaction ($\propto S_z(I_1^+I_2^-+I_1^-I_2^+)$). At all times $\delta=0$.
    
    The instantaneous central spin resets within the second stage of our protocol were effectuated by the following operation on the composite density operator:
    \begin{equation}
        \rho \to \ketbra{\downarrow_x} \otimes \Tr_{c} \rho,
    \end{equation}
    where $\Tr_c$ stands for the partial tracing with respect to the central spin degrees of freedom. 

    The steady state solver was applied to a continuous approximation of the protocol in which the central spin was reset continuously at an optical pumping rate $\Gamma_{\mathrm{op}}=2\pi/\tau_0$, where $\tau_0$ corresponds to the time between subsequent instantaneous central spin resets in the exact protocol. The collapse operator used in modelling that process was $\sqrt{\Gamma_{\mathrm{op}}}\ketbra{\downarrow_x}{\uparrow_x}$.

\subsection{Dynamics in a $I_1 \ne I_2$ manifold}\label{sec:fullquantumS}
    \begin{figure}[t!]
	\centering
		\includegraphics[width=0.5\textwidth]{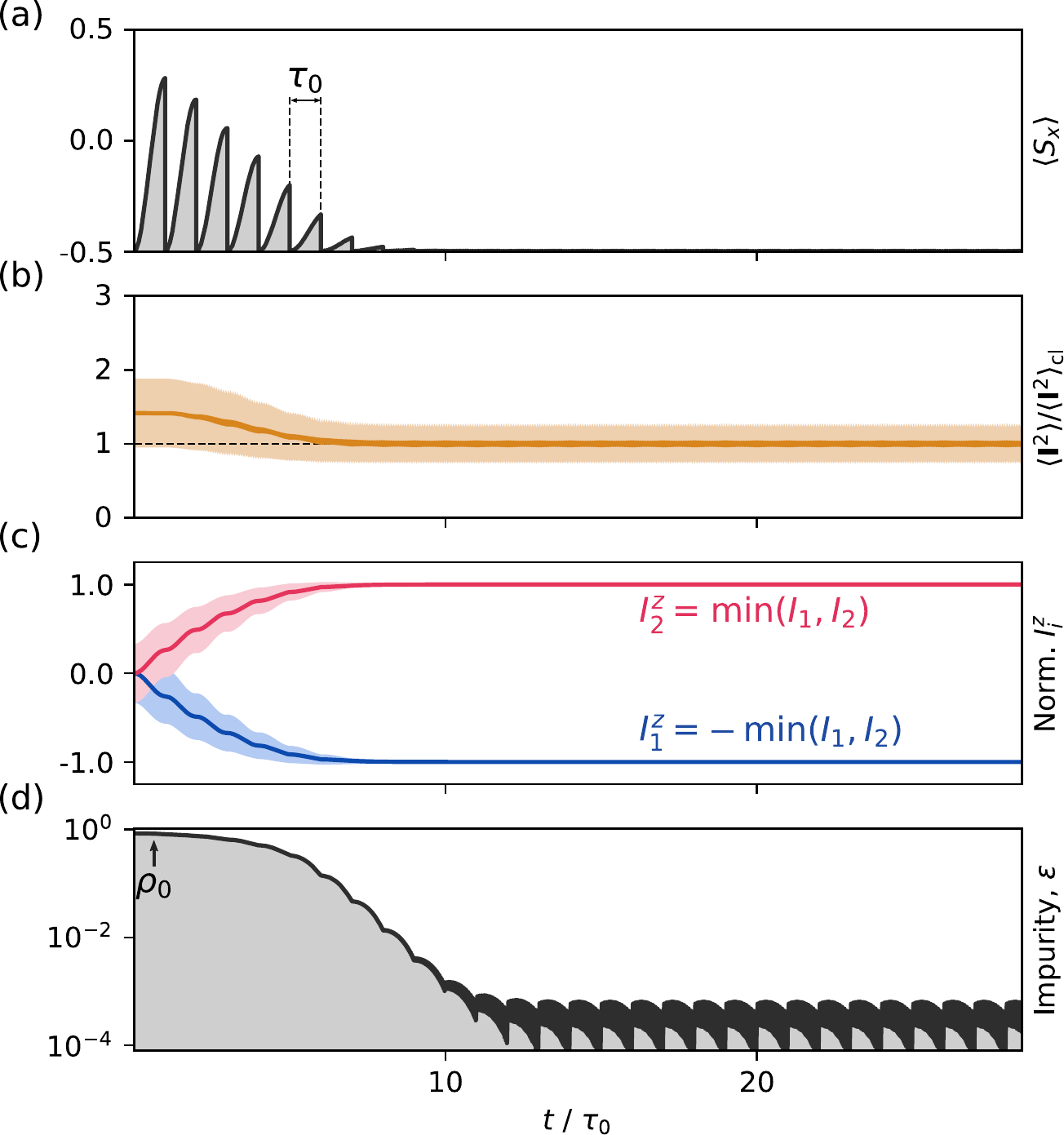}
		\caption{\textbf{Ideal system’s dynamics for $N = 32$, and $I_1 =\sqrt{32/2}-1$, $I_2 =\sqrt{32/2}+1$.
} Panels \textbf{a}-\textbf{d} are in a one-to-one correspondence to those of the Fig. 2 of the manuscript.} 
	\label{fig:SI_Unequal_spins}
	\end{figure}
 
We verify the protocol robustness in a $I_1\ne I_2$ manifold by performing an identical simulation to that from the manuscript section 'Ideal system dynamics'. Both $\kappa$ and $\tau_0$ are fixed to the same values, and the only difference lies in the choice of $I_1=\sqrt{32/2}-1$ and $I_2=\sqrt{32/2}+1$. The results are illustrated in the Fig.\ref{fig:SI_Unequal_spins}, and organized in a one-to-one correspondence with the Fig. 2 of the manuscript. As displayed in the Fig. \ref{fig:SI_Unequal_spins}a, the $\langle S_x \rangle$ population (that is, the only direct observable during the protocol) saturates to $-1/2$, like in the Fig. 2a of the manuscript. The polarizations $I_1^z$ and $I_2^z$ are driven towards the maximal possible opposite values, that is $\pm \min(I_1,I_2)$ (see Fig. \ref{fig:SI_Unequal_spins}c), as anticipated. Dynamics feature equal amount of purification after settling to the limit cycle (see Fig. \ref{fig:SI_Unequal_spins}d), and the only significant difference comes in the degree of $\langle \mathbf{I}^2\rangle$ reduction. Still, the protocol reaches the anticipated classical limit $\langle \mathbf{I}^2 \rangle_{\text{cl}}=(I_1+I_2)(|I_1-I_2|+1)$.
 
	\subsection{Equal impact of the ensemble and the central spin dephasing on the protocol's performance}

\begin{figure}[h!]
	\centering
		\includegraphics[width=0.5\textwidth]{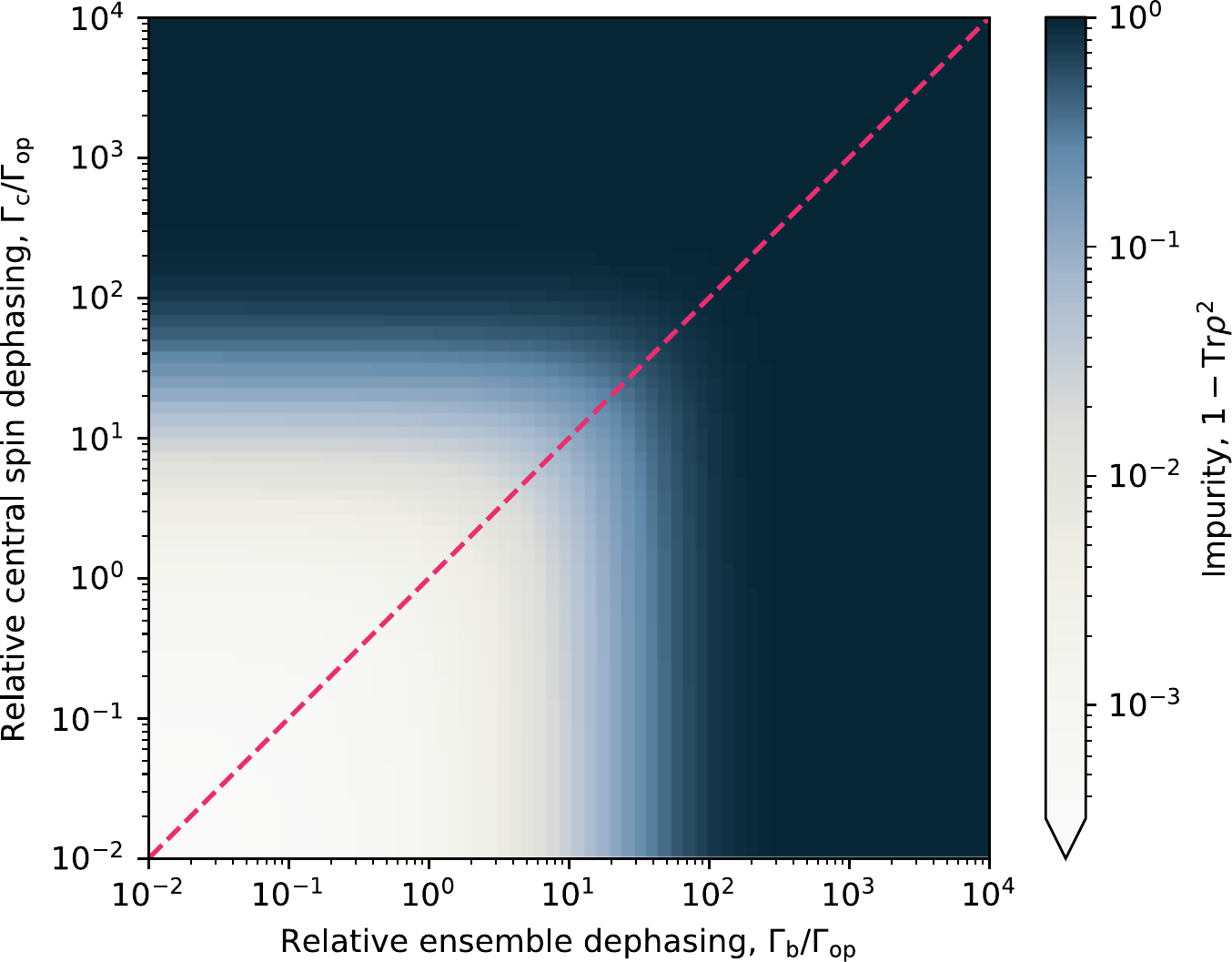}
		\caption{\textbf{Symmetry in the effect of ensemble and central spin dephasing on the protocol's performance.} The figure displays steady state impurity as a function of relative ensemble and central spin dephasing rates, $\Gamma_{\mathrm{b}}/\Gamma_{\mathrm{op}}$ and $\Gamma_{\mathrm{c}}/\Gamma_{\mathrm{op}}$, respectively. $\Gamma_{\mathrm{op}}$ stands for the optical pumping rate in continuous quantum model. The simulation was run for $\kappa=10$ and $N=200$ (i.e. $I_1=I_2=\sqrt{200/2}$).} 
	\label{fig:SI_Rel_deph}
	\end{figure}

    We studied the effects of dephasing of the central spin and the ensemble on the protocol's performance. The former process was modelled using a $\sqrt{\Gamma_{\mathrm{c}}}S_x$ collapse operator, and the latter made use of two distinct collapse operators: $\sqrt{\Gamma_{\mathrm{b}}/2}I_1^z$ and $\sqrt{\Gamma_{\mathrm{b}}/2}I_2^z$. Fig. \ref{fig:SI_Rel_deph} illustrates the steady state impurity calculated in the continuous protocol approximation for a range of dephasing rates $\Gamma_{\mathrm{c}}$ and $\Gamma_{\mathrm{b}}$. It is evident that both processes have quantitatively identical impact on the state preparation, which motivates introducing a total dephasing rate, $\Gamma_{\mathrm{d}}=\Gamma_{\mathrm{b}}+\Gamma_{\mathrm{c}}$, as a figure of merit in quantifying the resilience to system's imperfections. 

	\section{Rate equation model}
	
	\subsection{Evolution of populations and the scattering rates}

    A good quantitative understanding of the protocol dynamics can be gained from a simple rate equation model. Within this model, the dynamics is again restricted to the $I^z=0$ ladder of states, as a result of perfect total polarization locking. For convenience, we label the $n^{\text{th}}$ ladder state, $\ket{I_1^z=-(M-n),\,I_2^z=+(M-n)}$, with a principal quantum number: $\ket{n}$. The model assumes that at coarse-grained timescales, the coherences between different states across the ladder vanish, or in other words:
	\begin{equation}
		\Tr_{c}\rho = \sum_n p_n \ketbra{n}.
	\end{equation}
	
\begin{figure}[h!]
		\centering
		\includegraphics[width=0.5\textwidth]{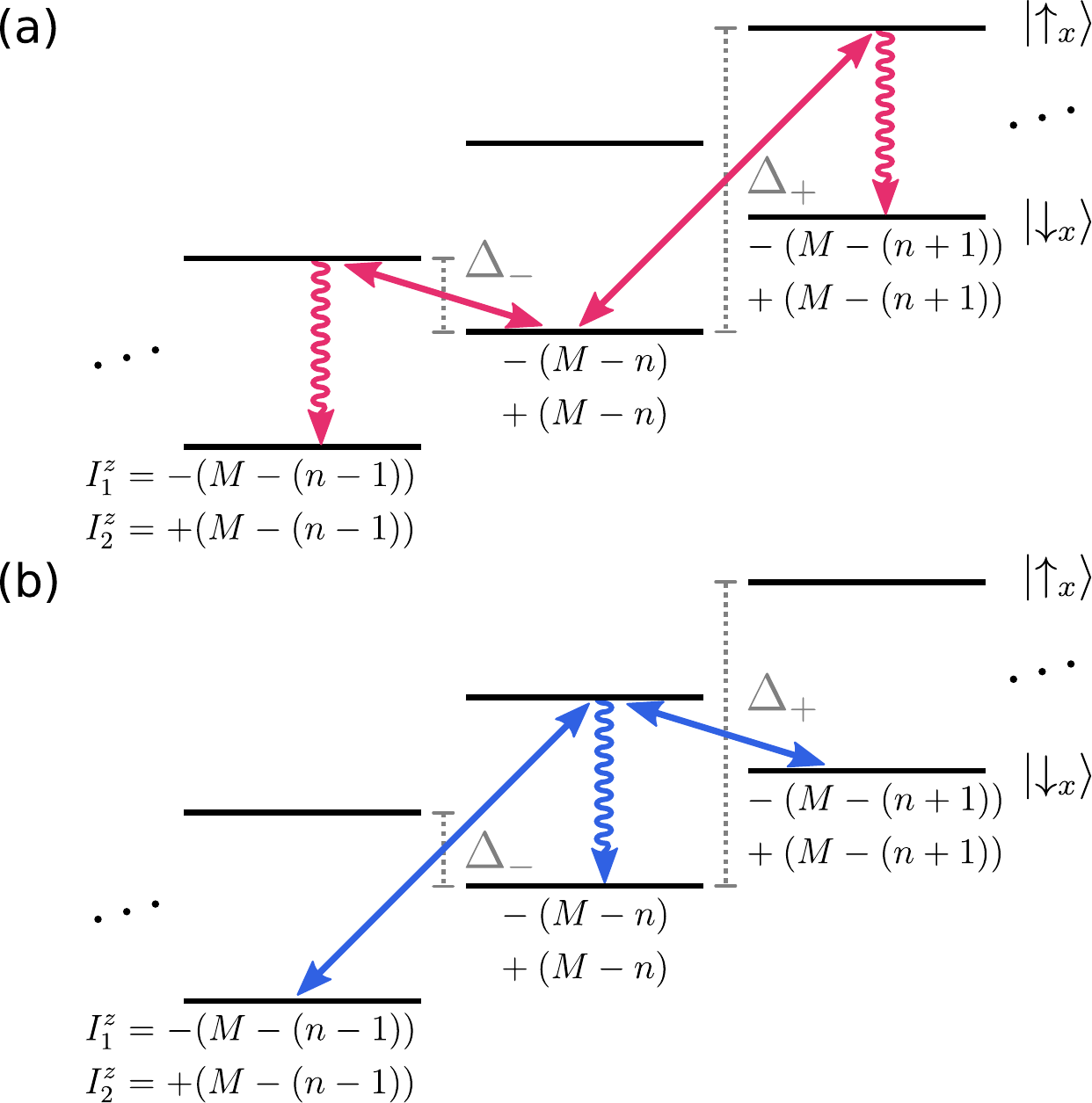}
		\caption{\textbf{Rate equation model of the protocol.}
			\textbf{a,} Scattering processes that depopulate $n^{\text{th}}$ state across the ladder. Straight arrows denote activated three-body interactions, whereas the gray bars show the energy gaps that suppress each of the processes. Wavy arrows illustrate the central spin reset.
			\textbf{b,} Scattering processes that populate $n^{\text{th}}$ state across the ladder. Labeling is consistent with that of the panel a.
		} 
		\label{fig:SI_RateEqn}
	\end{figure}

	The evolution of the population of the $\ket{n}$-state, $p_n$, is then captured by the following equation:
	\begin{equation}\label{eqn:rateqn}
		\Dot{p}_n=-(r^-_n+r^+_n)p_n + (r^+_{n-1} p_{n-1}+r^-_{n+1}p_{n+1}),    
	\end{equation}
    where $r_n^\pm$ stands for a rate of the population flow from $\ket{n}$-state to $\ket{n\pm1}$-state. The $\propto p_n$ term in the equation describes the 'out-flow' processes proceeding at a total rate $r^-_n+r^+_n$, as illustrated in the Fig. \ref{fig:SI_RateEqn}a. The first of the two processes involves a coherent exchange $\ket{\downarrow_x}\ket{n} \to \ket{\uparrow_x}\ket{n-1}$ driven by a three-body interaction, and detuned by $\Delta_-=\Omega - \Delta \omega$, followed by the central spin reset after time $\tau_0$ -- here approximated as proceeding concomitantly with the exchange, at the optical pumping rate $\Gamma_{\mathrm{op}} =\tfrac{2\pi}{\tau_0}$. The second process consists of the same two steps, except that the dynamics proceeds between  $\ket{n}$ and $\ket{n+1}$ states, for which the three-body interaction detuning is $\Delta_+=\Omega +\Delta \omega$. The rates of the outflow processes, $r_n^\pm$, are then proportional to the populations of the excited states $\ket{\uparrow_x}\ket{n\pm1}$, where the constants of proportionality are given by the central spin reset rate, $\Gamma_{\mathrm{op}}$. This yields:
	\begin{equation}\label{eqn:scatrates}
		r^\pm_n = \Gamma_{\mathrm{op}} \langle \ketbra{\uparrow_x, n\pm1} \rangle 
	\end{equation}
	We approximate the expectation value from Eq. \ref{eqn:scatrates} by a steady-state excited state population of a two level system constituted by $\ket{\downarrow_x, n}$ and $\ket{\uparrow_x, n\pm1}$ states; it follows that:
	\begin{equation}\label{eqn:scatrates2}
		r^\pm_n = \frac{\Gamma_{\mathrm{op}}}{2}  \frac{(\alpha^\pm_n)^2/(\Gamma_{\mathrm{op}} \Gamma^\prime)}{1+ (\alpha^\pm_n)^2/(\Gamma_{\mathrm{op}} \Gamma^\prime)+(\Delta_\pm/\Gamma^\prime)^2},
	\end{equation}
	where $\Gamma^\prime=\tfrac{1}{2}\Gamma_{\mathrm{op}}+\Gamma_{\mathrm{d}}$ incorporates the sum of phenomenological ensemble and central spin dephasing rates, $\Gamma_{\mathrm{d}}$.
	The effective drive strength, $\alpha^\pm_n$, corresponds to the three body interaction rate, given by:
	\begin{equation}
		\begin{split}
			\alpha_n^+&=\tfrac{a^2}{4\omega_{\text{c}}}(2I-n)(n+1), \\
			\alpha_n^-&= \tfrac{a^2}{4\omega_{\text{c}}}(2I-n+1)n. 
		\end{split}
	\end{equation}
	Its dependence on $I$ and $n$ is a result of collective enhancement, dependent on the total angular momenta of of sub-ensembles ($I_1=I_2=I \sim \sqrt{N/2}$).
	
	The second term in the equation \ref{eqn:rateqn} captures the effect of the 'in-flow' processes. The $\ket{n}$-state can be populated by $\Delta_+$-detuned process originating from $\ket{n-1}$-state, or a $\Delta_-$-detuned process originating from $\ket{n+1}$-state, as shown in the Fig. \ref{fig:SI_RateEqn}b. The relevant scattering rates are found using Eq. \ref{eqn:scatrates2}.
	
	\subsection{Steady state solution}
	
	The system reaches steady state when $\Dot{p}_n=0$ for all $n$. This generates the following recursive relation between the $\ket{n}$-state populations:
	\begin{equation}
		\begin{split}
			p_1&=\tfrac{r_0^+}{r_1^-}p_0, \\
			p_{n+2}&=\tfrac{r^-_{n+1}+r^+_{n+1}}{r^-_{n+2}}p_{n+1}-\tfrac{r^+_n}{r^-_{n+2}}p_n, \quad n=0,1,..,2I-2 .
		\end{split}
	\end{equation}
	This fully determines steady state populations, up to a normalization factor, which can be constrained using $\sum_n p_n=1$. 
	
	The computational complexity of solving for steady state populations is $\mathcal{O}(I)$; this is to be compared with $\mathcal{O}(I^6)$ complexity of solving for steady state of quantum master equation with dissipation.
	
	\subsection{Scaling of impurity, $\epsilon$, with directionality parameter, $\kappa$, and dephasing $\Gamma_{\mathrm{d}}\tau_0/(2\pi)$}
	
	The preparation performance can be characterized by the impurity of the reduced density operator, $\rho_\mathrm{b}=\Tr_c \rho$. The rate equation model 
	approximates this impurity by:
	\begin{equation}
		\epsilon=1-\Tr \rho_{\mathrm{b}}^2 = 1-\sum_n p_n^2. 
	\end{equation}
	
	In case of the perfect state preparation we have $p_0=1$, and $\epsilon=0$. The preparation performance starts to drop when the $\ket{1}$-state acquires a finite population, leading to an initial impurity increase (calculated at at $\Omega=\Delta \omega$):
 	\begin{equation}\label{eqn:reduced_purity}
		\epsilon \approx 2 \frac{r_0^+}{r_1^-}= 2\Bigg[ 1+\frac{4(\Delta \omega)^2}{\Gamma_{\text{op}}\Gamma^\prime +\alpha^2}\frac{\Gamma_{\text{op}}}{\Gamma^\prime}\Bigg]^{-1},
	\end{equation}
        where we introduced $\alpha\equiv\alpha_0^+=\alpha_1^-$. Since $\alpha^2 \propto N$, and $\Gamma_{\text{op}}\Gamma^\prime \propto N^2$ (as $\Gamma_{\text{op}}=Na^2/(2\omega_{\text{c}})$, for the optimal value of $\tau_0$), in the thermodynamic limit, this expression simplifies further:
        \begin{equation}
        \epsilon \approx \frac{2}{1+64\kappa^2(1+2\Gamma_{\mathrm{d}}/\Gamma_{\text{op}})^{-2}},
        \end{equation}     
    where $\kappa$:
	\begin{equation}
		\kappa = \frac{\omega_{\text{c}} \Delta \omega}{Na^2},
	\end{equation}
    as defined in the manuscript. The requirement for the cooling to proceed optimally is therefore $\kappa \gtrsim 1$ and $\Gamma_{\text{d}}/\Gamma_{\text{op}} \lesssim 1$.
	
	\subsection{Convergence Time, $T_c$}
	
	The rate equation formalism allows to estimate the convergence time of the protocol, $T_c$, from spectral analysis of the matrix $\Lambda$, that generates the rate equation as:
	\begin{equation}
		\dot{\mathbf{p}} = \Lambda \mathbf{p}
	\end{equation}
	where $\mathbf{p}$ is a vector of $\ket{n}$-state populations.
	
	We notice that due to the contractive nature of dynamics, the real parts of the $\Lambda$-matrix eigenvalues are all non-positive, and can be arranged according to: \begin{equation}
		0= \lambda_0 > \Re(\lambda_1) > \Re{\lambda_2} >... 
	\end{equation} 
	We then identify $T_c=2\pi / |\Re(\lambda_1)|$, as  $|\Re(\lambda_1)|$ is the slowest convergence rate in the model.
\newpage
\section{Tables of physical parameters for candidate systems}

\begin{table}[h!]
\begin{tabular}{|c|l|c|}
\hline
Total hyperfine interaction, $A$                &  & $2\pi \times 11$ \,\text{GHz}                    \\ \hline
Zeeman frequency difference, $\Delta \omega/B$ &  & $2\pi \times5.76 $\,\text{MHz/T }                 \\ \hline
Central spin splitting,  $\omega_{\text{c}}/B$                      &  & $2\pi \times1.3$\,\text{GHz/T}                 \\ \hline
Effective number of nuclei, $N$             &  & $ 10^5$               \\ \hline
Dephasing rate (nuclear), $\Gamma_d$               &  & $2\pi \times 10$ kHz \\ \hline
Magnetic field ranges, $B$                                &  & 1T - 10T                  \\ \hline
\end{tabular}
\caption{Model parameters used to benchmark the protocol's performance for GaAs-AlGaAs QDs \cite{Zaporski2022, Zhai2020, PhysRevB.97.235311, PhysRevLett.118.177702}.}
\bigskip
\bigskip

\begin{tabular}{|c|l|c|}
\hline
Total hyperfine interaction, $A$                &  & $2\pi \times 11$ \,\text{GHz}                    \\ \hline
Zeeman frequency difference, $\Delta \omega/B$ &  &  $2\pi \times 5.76$ \, \text{MHz/T}                  \\ \hline
Central spin splitting, $\omega_{\text{c}}/B$                     &  & $2\pi \times 6$ \, \text{GHz/T}                   \\ \hline
Effective number of nuclei, $N$             &  & $ 10^5$               \\ \hline
Dephasing rate (nuclear), $\Gamma_d$                              &  & $2\pi \times 10$ MHz \\ \hline
Magnetic field ranges, $B$             &  & 100mT - 10T               \\ \hline
\end{tabular}
\caption{Model parameters used to benchmark the protocol's performance for InGaAs-GaAs QDs \cite{Stockill2016,PhysRevLett.118.177702}.}
\bigskip
\bigskip

\begin{tabular}{|c|l|c|}
\hline
Total hyperfine interaction, $A$               &  & $2\pi\times11$\,\text{GHz}                    \\ \hline
Zeeman frequency difference, $\Delta \omega/B$ &  &  $2\pi \times 5.76$ \,\text{MHz/T}                  \\ \hline
Central spin splitting, $\omega_{\text{c}}/B$               &  & $2\pi \times 8$ \,\text{GHz/T}                   \\ \hline
Effective number of nuclei, $N$                &  & $ 10^6$               \\ \hline
Dephasing rate (nuclear), $\Gamma_d$               &  & $ 2\pi \times 10$ \text{kHz} \\ \hline
Magnetic field ranges, $B$             &  & 100mT - 10T               \\ \hline
\end{tabular}
\caption{Model parameters used to benchmark the protocol's performance for Gate Defined GaAs QDs \cite{Bluhm2011,PhysRevB.97.235311, PhysRevLett.118.177702}.}
\bigskip
\bigskip

\begin{tabular}{|c|c|c|}
\hline
Three-body interaction strength, $a_i a_j/\omega_{\text{c}}$ &                       & $ 2\pi \times 0.1$ kHz            \\ \hline
Quadrupolar frequency difference, $\Delta \omega$        &                       & $2\pi \times$10-100 kHz          \\ \hline
Number of nuclei                                    & \multicolumn{1}{l|}{} & 4                                \\ \hline
Dephasing rate, $\Gamma_d$                          &                       & $2\pi \times 1.25$ kHz \\ \hline
\end{tabular}
\caption{Model parameters used to benchmark the protocol's performance for a ${}^{171}\text{Yb}^{3+}:\text{YVO}_4$ (REI)\cite{Ruskuc2022}. The three-body interaction in this system is a central spin mediated second-order dipole-dipole, and it couples the nuclei in the second shell (subscript $i$) with those in the outer (subscript $j$) shells with an effective strength $\sim a_{i}a_{j}/\omega_{\text{c}}$. The $\Delta \omega$ corresponds to the difference in quadrupolar frequencies of the nuclei in the second shell (i.e. a frozen core) and the outer shells.}
\end{table}

	\section{Optimizing preparation of many-body singlet}
	\subsection{Central spin gate}
	We introduce the following notation for a single qubit gate:
	\begin{equation}
		(\phi)_j \equiv \exp{-i\phi \sigma_j/2}, \quad j=x,y,z
	\end{equation}
	where $\sigma_x$, $\sigma_y$, and $\sigma_z$ are $2\times2$ Pauli matrices.
	
	\subsection{Three-body interaction gate}
	
	Upon matching the $\Omega=\Delta \omega$ resonance condition, the evolution of the system in the frame rotating with $\tfrac{1}{2}\Delta\omega(I_1^z-I_2^2)+\Omega S_x$ is dictated by the following Hamiltonian:
	\begin{equation}
		H_{\mathrm{exc}}=\tfrac{a^2}{4\omega_{\text{c}}}(\ketbra{\uparrow_x}{\downarrow_x} \mathcal{I}_- + \ketbra{\downarrow_x}{\uparrow_x}\mathcal{I}_+),
	\end{equation}
	where $\mathcal{I}_\pm=I_1^\pm I_2^\mp$ are the non-linear ladder operators for the effective Jaynes-Cummings ladder of $\ket{n}$-states, whose action is captured by:
	\begin{equation}\label{Eq:enhfct}
		\mathcal{I}_+ =\sum_{n=0}^{2I-1}\underbrace{(2I-n)(n+1)}_{e_n}\ketbra{n+1}{n},
	\end{equation}
	and $\mathcal{I}_-=(\mathcal{I}_+)^\dagger$. 
	
	Evolution of the system in that frame over time $\tau$, generated by $H_{\mathrm{exc}}$, realises the following gate:
	\begin{equation}\label{eqn:true_gate}
		\begin{split}
			&\exp{-i \tau H_{\mathrm{exc}}}=\\
			&\mathds{1}+ \ketbra{\uparrow_x}\otimes \sum_{n=0}^{2I-1}\Big(\cos \frac{a^2e_n \tau}{4\omega_{\text{c}}} -1 \Big) \ketbra{n} \\
			&+\ketbra{\downarrow_x}\otimes\sum_{n=0}^{2I-1}\Big(\cos \frac{a^2e_n \tau}{4\omega_{\text{c}}} -1\Big) \ketbra{n+1} \\
			&-\ketbra{\uparrow_x}{\downarrow_x}\otimes\sum_{n=0}^{2I-1} i \sin \frac{a^2e_n \tau}{4\omega_{\text{c}}} \ketbra{n}{n+1} \\
			&-\ketbra{\downarrow_x}{\uparrow_x}\otimes\sum_{n=0}^{2I-1} i \sin \frac{a^2e_n \tau}{4\omega_{\text{c}}} \ketbra{n+1}{n}. \\
		\end{split}
	\end{equation}
	In particular, it is readily seen that the exact $\pi$-gate time, $\tau_\pi$, is dependent on $n$ since the enhancement factors $e_n$ vary across the ladder of states.
	
	In contrast, for the simplified system with $e_n = \mathrm{const}$, the exchange interaction $\pi$-gate would be: 
	\begin{equation}\label{eqn:approx_gate}
		\begin{split}
			&\exp{-i\tau_\pi H_{\mathrm{exc}}^0} = \ketbra{\downarrow_x}\otimes\ketbra{0}+ \ketbra{\uparrow_x}\otimes\ketbra{2I} \\
			&-i\ketbra{\uparrow_x}{\downarrow_x}\sum_{n=0}^{2I-1} \ketbra{n}{n+1}-i\ketbra{\downarrow_x}{\uparrow_x}\sum_{n=0}^{2I-1} \ketbra{n+1}{n}.
		\end{split}
	\end{equation}

    \subsection{Optimizing the gates via gradient descent}
            \begin{figure}[t!]
		\centering
		\includegraphics[width=0.46\textwidth]{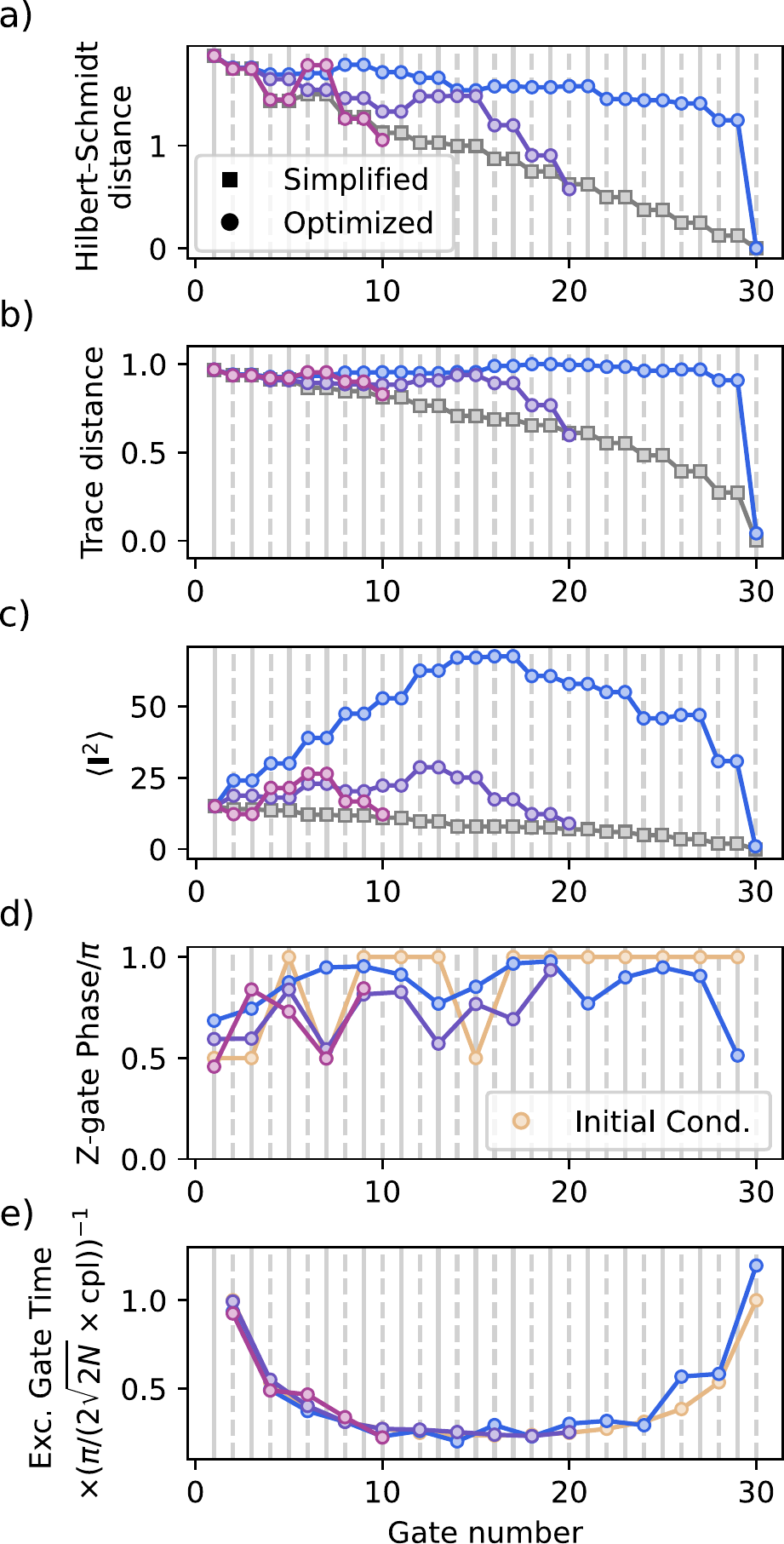}
		\caption[width=\textwidth]{\textbf{Optimizing gate times and phases for singlet preparation.} \textbf{a}, Hilbert-Schmidt distance (i.e. our cost function, $\mathcal{C}$) during simplified (squares) and optimized (circles) protocols of varied length. The colorcoding is consistent with Fig.6c of the manuscript, which only concerns the protocols' endpoints.
      \textbf{b}, Evolution of the trace distance in considered protocols. \textbf{c}, Evolution of the expectation value of total angular momentum squared in considered protocols. \textbf{d}, Optimal Z-gate phases in protocols of varied length. The beige circles correspond to initial condition for the gradient descent optimization.
      \textbf{e}, Optimal exchange gate times in protocols of varied length. The beige circles correspond to initial condition for the gradient descent optimization. `cpl' in the label stands for coupling strength, $a^2/(4\omega_{\text{c}})$.} 
		\label{fig:SI_GRAPE}
	\end{figure}
 
    In order to tailor the sequence parameters in a real system, we minimize the appropriately constructed cost function, which takes the following quantum state:
    \begin{equation}\label{ansatz}
        \ket{\psi}= \prod^{\xleftarrow[]{}}_{j=0} \Big(\exp{-i \tau_j H_{\mathrm{exc}}} \exp{-i\phi_j \sigma_z/2} \Big) \ket{\downarrow_x}\otimes \ket{n=0},
    \end{equation}
    as an input (a method originally devised in Ref. \cite{KHANEJA2005296}). The arrow over the product sign represents direction of stacking the consecutive terms with $j=0,1,2,..$ in the product. The minimization is done with respect to the variational parameters $\{\tau_j,\phi_j\}_{j=0,1,2,..}$, which correspond directly to the three-body interaction activation times ($\tau_j$), and phases of central spin gates ($\phi_j$). To reach the minimum of the cost function we apply the RMSprop gradient descent algorithm in the space of $\{\tau_j,\phi_j\}_{j=0,1,2,..}$. 
    
    \subsubsection{Choice of the cost function}
    
    We choose a Hilbert-Schmidt distance between the singlet state, $\chi=\ketbra{I=0}$, and the reduced density operator of the ensemble, $\rho_{\mathrm{b}}=\Tr_c(\ketbra{\psi})$, as our cost function:
    \begin{equation}\label{eqn:cost_function}
        \mathcal{C}(\ket{\psi})=\Tr (\rho_{\mathrm{b}}-\chi)^\dagger(\rho_{\mathrm{b}}-\chi).
    \end{equation}
    The main advantage of working with this cost function is an ease of calculating its gradient analytically, which speeds up the optimization procedure. To see it, we first act with the differential operator, on the $\ket{\psi}$-state from the Eq. \ref{ansatz}, and find:
    \begin{equation}
    \begin{split}
\partial_{v_i}\ket{\psi}&=\Big(\prod^{\xleftarrow{}}_{j>i}e^{-iv_jX_j}\Big)\Big(-iX_i e^{-iv_iX_i}\Big) \\
&\times \Big(\prod^{\xleftarrow{}}_{j<i}e^{-iv_jX_j}\Big)\ket{\downarrow_x}\otimes\ket{n=0},
    \end{split}
    \end{equation}
    for variational parameters $v_j\in\{\tau_j, \phi_j\}$ and operators $X_j\in \{H_{\mathrm{exc}},\sigma_z/2\}$. We then notice that extending this result to the cost function from Eq. \ref{eqn:cost_function} amounts to a simple application of a chain rule.

    
    

    \subsubsection{Hyperparameters and convergence of an RMSprop algorithm}
    
    The update rule for a given variational parameter, $v$, in our implementation of the RMSprop algorithm is:
    \begin{equation}
    \begin{split}
        v_{i+1}=v_{i}-\frac{\zeta}{\sqrt{\xi +\mathbb{E}_{i}[(\partial_v \mathcal{C})^2]}}\times &[\partial_v \mathcal{C}]_i\\
        \mathbb{E}_{i}[(\partial_v \mathcal{C})^2]=\beta\mathbb{E}_{i-1}[(\partial_v \mathcal{C})^2] + (1-\beta)&[\partial_v \mathcal{C}]_i^2
    \end{split}
    \end{equation}
    where $\partial_v \mathcal{C}$ is the partial derivative of the cost function, $\mathcal{C}$, with respect to the variational parameter, $v$. The hyperparameters used throughout the optimization routines were $\zeta=0.015$, $\xi=10^{-8}$, and $\beta=0.85$.

    For all the sequences of lengths shorter than the maximal considered, the convergence was reached in less than $1000$ optimization epochs. For the longest considered sequence, optimization was terminated after $7000$ epochs, however, the cost function could likely be decreased further. The difficulty of this task comes most likely from a presence of a barren plateau in a cost function landscape - i.e. region around the minimum, where the gradients of $\mathcal{C}$ vanish exponentially fast\cite{Cerezo2021}. 
    
    The evolutions of the cost function, $\mathcal{C}$, the trace distance, and the expectation value of total angular momentum squared during the optimal protocols of varied length, are plotted in the Fig. \ref{fig:SI_GRAPE}a, Fig. \ref{fig:SI_GRAPE}b, Fig. \ref{fig:SI_GRAPE}c, respectively (circles). Their simplified system's counterparts are plotted alongside, for reference (squares). 
    
    \subsubsection{Initial condition for gradient descent}
    
    To speed up the convergence of the gradient descent optimization, we start the procedure with a physically-motivated initial guess for variational parameters. 
    
    The initial $z$-gate phases are chosen to be identical to those in the optimal protocol for the idealized system (see the main text, and the beige circles in the Fig. \ref{fig:SI_GRAPE}d).  
    
    The initial activation times for the three body interaction are chosen as:
    \begin{equation}
    \tau_{j}=\pi/\Big(2e_j\times \frac{a^2}{4\omega_{\text{c}}}\Big)    
    \end{equation}
    where $e_j$ stands for the enhancement factor from the Eq. \ref{Eq:enhfct} - see the beige circles in the Fig. \ref{fig:SI_GRAPE}e. The intuition behind this choice, is that it realizes the ideal amplitude transfer for each consecutive $\ket{n}$-state brought into the superposition.

\subsection{Singlet auto-refocusing in a non-rotating frame}

Singlet state prepared in the frame rotating with $\tfrac{1}{2}\Delta\omega(I_1^z-I_2^z)+\Omega S_x$,
will coincide with a singlet state in a non-rotating frame at times $t=2\pi k /\Delta\omega$ for $k=0,1,2,3..$; indeed in a non-rotating frame the evolution of the prepared state (up to a global phase) is given by: 
\begin{equation}
 \ket{\psi(t)}= \ket{\downarrow_x}\otimes \sum_{n=0}^{2I}\frac{(-1)^n}{\sqrt{2I+1}} \exp{itn\Delta \omega }\ket{n}.
\end{equation}
 

\newpage
	\newpage
	\bibliography{singlet}